\documentclass[a4paper, 12pt]{article}
\usepackage{epsfig}
\usepackage{amssymb}
\usepackage{amsmath}
\usepackage{titelblatt}

\renewcommand{\theequation}{\mbox{\arabic{section}.\arabic{equation}}}
\newtheorem{theorem}{Theorem}[section] 
\newtheorem{lemma}{Lemma}[section]
\newtheorem{cor}{Corollary}[section]
\newtheorem{remark}{Remark}[section]

\numberwithin{equation}{section}

\newcommand{\D}{{\rm d}}
\newcommand{\dx}{\, \D x}

\newcommand{\dr}{\, \D r}
\newcommand{\dt}{\, \D t}

\newcommand{\dth}{\, \D \theta}

\newcommand{\wnull}[2]{\stackrel{\circ}{W}\!\! {}^{#1}_{#2}}
\newcommand{\wnulltext}[2]{\stackrel{\circ}{W}\!\!\! {}^{#1}_{#2}}

\newcommand{\rz}{\mathbb{R}}

\newcommand{\loc}{\operatorname{loc}}
\newcommand{\spt}{\operatorname{spt}}

\newcommand{\newmathop}[2]{\newcommand{#1}{\mathop{\mit{#2}}}}

\newmathop{\Minttext}{\int\limits\hspace{ -0.92em}-}
\newmathop{\Mint}{\int\hspace{ -1.05em}-}

\newcommand{\reff}[1]{(\ref{#1})}

\renewcommand{\theequation}{\mbox{\arabic{section}.\arabic{equation}}}
\newcounter{formel}

\addtocounter{section}{0}

\title{\bf Liouville-type theorems for steady flows of degenerate power law fluids in the plane}
\author{Michael Bildhauer \and Martin Fuchs \and Guo Zhang}
\date{\today}

\begin{document}


\Panzahl    {2}                   
\Pautor     {Michael Bildhauer}       

\Panschrift {Saarland University \\ Department of Mathematics \\
             P.O. Box 15 11 50   \\ 66041 Saarbr\"ucken \\
             Germany}
\Pepost     {bibi@math.uni-sb.de}     

\Ptitel     {Liouville-type theorems for steady flows of degenerate power law fluides in the plane}  

\Pjahr      {2012}               
\Pnummer    {}                 

\Pdatum     {\today}             

\Pcoautor   {Martin Fuchs}   
\Pcoanschrift {Saarland University \\ Department of Mathematics \\
             P.O. Box 15 11 50   \\ 66041 Saarbr\"ucken \\
             Germany}
\Pcoepost   {fuchs@math.uni-sb.de}     

\qPautor     {}     
\qPanschrift { }
\qPepost     {}   

\qPcoautor   {}        
\qPcoanschrift { }
\qPcoepost   {}     

\qqPautor   {}      
\qqPanschrift { }
\qqPepost   {}  



\renewcommand{\div}{\operatorname{div}}

\newcommand{\iom}{\int_{\Omega}}

\newcommand{\ib}[1]{\int_{B_{#1}(x_0)}}
\newcommand{\ibo}[1]{\int_{B_{#1}(0)}}

\newcommand{\ia}[1]{\int_{T_{#1}(x_0)}}
\newcommand{\iao}[1]{\int_{T_{#1}(0)}}

\newcommand{\da}{\Delta^\alpha_h}

\newcommand{\para}{\kappa}

\newcommand{\msp}{\;\;}
\newcommand{\fsp}{\quad\;}
\newcommand{\psp}{\;}
\newcommand{\arsp}{\hspace*{\arraycolsep}}
\newcommand{\dis}{\displaystyle}
\newcommand{\eps}{\varepsilon}

\parindent2ex
\maketitle

\renewcommand{\thefootnote}{}
\begin{abstract}
We extend the Liouville-type theorems of Gilbarg and Weinberger and of Koch, Nadirashvili, Seregin and Sver\'{a}k
valid for the stationary variant of the classical Navier-Stokes equations in $2D$ to the degenerate power
law fluid model.
\end{abstract}

\section{Introduction}\label{in}
To begin with we look at a velocity field $u$: $\rz^2\to \rz^2$ and a pressure function $\pi$: $\rz^2 \to \rz$ 
satisfying the stationary equations of Navier-Stokes
\begin{equation}\label{in 1}
\left. \begin{array}{rcl}
\dis - \Delta u + u^k\partial_k u + \nabla \pi &=& 0  \psp ,\\
\div u &=& 0\fsp\mbox{on}\msp \rz^2\psp ,
\end{array}\right\}
\end{equation}
which correspond to the flow of an incompressible Newtonian fluid with constant viscosity
(w.l.o.g.~equal to $1$). Here we study entire solutions, and a natural question is the search
for suitable conditions which force $u$ (and thereby $\pi$) to be constant. 
We recall two prominent examples
of such Liouville-type results for the Navier-Stokes equation \reff{in 1}: if $u$ is a finite energy solution,
i.e.~if we have
\begin{equation}\label{in 2}
\int_{\rz^2} |\nabla u|^2 \dx < \infty \psp ,
\end{equation}
then Gilbarg and Weinberger \cite{gw} proved $u = const$ making extensive use of the fact that
the vorticity function $\omega := \partial_2 u^1 - \partial_1 u^2$ satisfies a nice elliptic
equation. Recently, Koch, Nadirashvili, Seregin and Sver\'{a}k \cite{knss} discussed the instationary variant
of \reff{in 1} and, as a byproduct of their investigations, they showed that 
in the stationary case \reff{in 2} can be replaced
by
\begin{equation}\label{in 3}
\sup_{x\in \rz^2} |u(x)| < \infty
\end{equation}
implying the constancy of the vector field $u$. In connection with the Navier-Stokes equation we
like to remark that according to \cite{zh} the hypothesis
\[
\int_{\rz^2} |u|^t \dx < \infty \fsp\mbox{for some}\msp t > 1
\]
(replacing \reff{in 1} or \reff{in 3}) implies the vanishing of $u$, whereas in \cite{fzho} it is observed
that $u=const$ is still true if the growth of $|u(x)|$ as $|x|\to \infty$ is not too strong.\\

In  \cite{fu}, \cite{fzha}, \cite{zh} the situation for generalized
Newtonian fluids being either of shear thickening or shear thinning type is studied. 
For this case equation
\reff{in 1} has to be replaced by
\begin{equation}\label{in 4}
\left. \begin{array}{rcl}
\dis - \div\big[DH(\eps(u))\big]  + u^k\partial_k u + \nabla \pi &=& 0  \psp ,\\
\div u &=& 0\fsp\mbox{on}\msp \rz^2
\end{array}\right\}
\end{equation}
with a strictly convex potential $H$ of class $C^2$ acting on symmetric ($2\times 2$)-matrices
($\eps(u)$ denoting the symmetric gradient of the velocity field $u$) and being of the form
\begin{equation}\label{in 5}
H(\eps) = h(|\eps|)
\end{equation}
for a function $h$: $[0,\infty) \to [0,\infty)$ for which 
\[
\mu(t) := \frac{h'(t)}{t}
\]
either decreases or increases. Note that according to \reff {in 5} we have 
$DH(\eps) = \mu(|\eps|)\eps$, thus $\mu$ plays the role of a shear dependent viscosity.
For further physical and mathematical explanations we refer to the monographs
\cite{la}, \cite{ga}, \cite{ga2}, \cite{mnrr} or \cite{fs}.

The most severe restriction concerns the existence and the behaviour of $D^2H(0)$,
which in particular means that we require
\begin{equation}\label{in 6}
D^2H(0)(\eps,\eps) \geq \lambda |\eps|^2
\end{equation}
for some positive constant $\lambda$. Assuming \reff{in 6} it is shown:
suppose that $u \in C^1(\rz^2,\rz^2)$ is an entire weak solution of \reff{in 4}, i.e.~it holds
$\div u =0$ together with
\begin{equation}\label{in 7}
0 = \int_{\rz^2} DH(\eps(u)):\eps(\varphi) \dx + \int_{\rz^2} u^k \partial_k u^i \varphi^i \dx
\end{equation}
for all $\varphi \in C^\infty_0(\rz^2,\rz^2)$ such that $\div \varphi = 0$.
Then we have $u\equiv const$, if either \reff{in 3} holds or if we replace \reff{in 2} through the
appropriate hypothesis
\begin{equation}\label{in 8}
\int_{\rz^2} h(|\nabla u|) \dx < \infty \psp .
\end{equation}

Clearly these results apply to non-degenerate $p$-fluids for which
$h(t) = (1+t^2)^{p/2}$ (modulo physical constants) with exponent $p\in (1,\infty)$
but not to the degenerate power law model, i.e.~to the potential $H$ with function
$h(t) = t^p$.\\

In the present paper we are going to investigate the degenerate $p$-case, i.e.~from now on we
assume that $H$ is given by
\[
H(\eps) = |\eps|^p
\]
for some $1 < p < \infty$ and that $u\in C^1(\rz^2,\rz^2)$ with $\div u =0$ solves
equation \reff{in 7}.
Then our results are as follows:

\begin{theorem}\label{theo 1}
Suppose that $1 < p \leq 2$. 
\begin{enumerate}
\item If $u$ belongs to the space $L^\infty(\rz^2,\rz^2)$,
i.e.~if condition \reff{in 3} holds, then $u$ is a constant vector.
\item If $p < 2$, if
\begin{equation}\label{theo 1 1}
0 < \alpha < \frac{2-p}{6+p}
\end{equation}
and if we have 
\begin{equation}\label{theo 1 2}
\limsup_{|x|\to \infty}  |u(x)| |x|^{-\alpha}< \infty \psp ,
\end{equation}
then the conclusion of $i$) holds.
\end{enumerate}
\end{theorem}

\begin{remark}\label{in rem 1}
For the choice $p=2$ we reproduce the contribution of Koch, Nadirashvili, Seregin and Sver\'{a}k \cite{knss},
for $1 < p < 2$ condition \reff{theo 1 2} allows even a certain growth of $|u(x)|$ as $|x|\to \infty$.
In Theorem \ref{theo 5} we will discuss in more detail the admissible a priori growth rates of $u$ in
the case $p=2$.
\end{remark}

The next two theorems extend the Liouville result of Gilbarg and Weinberger \cite{gw} to exponents $p$
not necessarily equal to $2$.

\begin{theorem}\label{theo 2}
Let $6/5 < p \leq 2$ and assume that 
\[
\int_{\rz^2} |\nabla u|^p \dx < \infty\psp ,
\]
which means that \reff{in 8} is satisfied. Then $u$ has to be constant.
\end{theorem}

\begin{theorem}\label{theo 3}
Theorem \ref{theo 2} remains valid for exponents $p\in [2,3]$.
\end{theorem}

Theorem \ref{theo 4} is the counterpart to Theorem \ref{theo 1}, $ii$) for $p > 2$ 
involving formally the same exponent $(p-2)/(p+6)$.

\begin{theorem}\label{theo 4}
Let $p>2$ and let $u_\infty \in \rz^2$ denote a vector such that
\begin{enumerate}
\item in case $2 < p < 6$
\begin{equation}\label{theo 4 1}
\sup_{|x| \geq R} |u(x)-u_\infty| |x|^{\frac{p-2}{p+6}} \to 0 \fsp \mbox{as}\msp R\to \infty \psp ;
\end{equation}
\item in case $p=6$:
\begin{equation}\label{theo 4 2}
\limsup_{|x|\to \infty}  |u(x) - u_\infty| |x|^{\frac{1}{3}} < \infty \psp ;
\end{equation}
\item in case $p >6$:
\begin{equation}\label{theo 4 3}
\sup_{|x| \geq R}  |u(x) - u_\infty| |x|^{\frac{1}{3}} \to 0 \fsp\mbox{as}\msp R \to \infty \psp .
\end{equation}
\end{enumerate}
Then $u\equiv u_{\infty}$ follows.
\end{theorem}

\begin{remark}\label{in rem 2}
It remains an open question, if in case $p >2$ bounded solutions are constant
without imposing a decay condition.
\end{remark}

An inspection of the proofs of Theorem \ref{theo 1} - \ref{theo 4} will show:

\begin{cor}\label{in cor 1}
Let $p\in (1,\infty)$ and suppose that $u$: $\rz^2 \to \rz^2$ is a solution of the $p$-Stokes system
in the plane, i.e.~a solution of \reff{in 7} with $H(\eps) = |\eps|^p$, where now the convective term
is neglected. Then $u$ is a constant vector if either $u\in L^\infty(\rz^2,\rz^2)$ or if $u$ is of finite energy,
i.e.~$\int_{\rz^2} |\nabla u|^p \dx < \infty$.
\end{cor}

\begin{remark}\label{in rem 3}
Clearly Corollary \ref{in cor 1} can be generalized in the sense that for $1 < p < 2$ a certain growth
of $u$ can be included which might be even stronger in comparison to the
formulation given in \reff{theo 1 1} and \reff{theo 1 2}. We leave the details to the reader.
\end{remark}

We finish this introduction with an extension of the Liouville results obtained in \cite{knss}
and \cite{fzho} for the case of the classical Navier-Stokes equation.

\begin{theorem}\label{theo 5}
Suppose that $u$: $\rz^2 \to \rz^2$ is a solution of \reff{in 1} such that
\begin{equation}\label{theo 5 1}
\limsup_{|x|\to \infty} |u(x)| |x|^{-\alpha} < \infty
\end{equation}
for some $\alpha < 1/3$. Then the constancy of $u$ follows.
\end{theorem}

\begin{remark}\label{in rem 4}
It would be interesting to know the optimal bound for the number $\alpha$ occurring in \reff{theo 5 1}.
\end{remark}

Our paper is organized as follows: in Section \ref{pre} we give estimates for the energy
$\ib{r} |\nabla u|^p \dx$, $1 < p < \infty$, on disks in terms of the radius under various hypotheses
imposed on $u$. Section \ref{sub} is devoted to the case $1 < p < 2$, i.e.~we will present
the proofs of Theorem \ref{theo 1} and of Theorem \ref{theo 2} by combining the results
of Section \ref{pre} with estimates for the ``second derivatives'' due to Wolf \cite{wo}.

Since these estimates are not available for $p > 2$, we have to find alternatives leading to
Theorem \ref{theo 3} and to Theorem \ref{theo 4}. This is done in Section \ref{sup}.

In Section \ref{quad} we give a proof of Theorem \ref{theo 5}.
Moreover, we collect some technical tools in an appendix.\\

Acknowledgement: We thank J\"org Wolf for valuable discussions.

\section{Estimates for the $p$-energy on disks}\label{pre}

In this section we describe the growth of the energy $\ib{r} |\nabla u|^p \dx$
of weak solutions $u$ to \reff{in 7} in terms of the radius of the disk under various
conditions concerning the growth of $u$.

\begin{lemma}\label{pre lem 1}
Let $u\in C^1(\rz^2,\rz^2)$, $\div u =0$, denote a solution of \reff{in 7} for the choice
$H(\eps) = |\eps|^p$ with exponent $p\in (1,\infty)$. 
\begin{enumerate}
\item Then, for any real number $\beta< 1$, it holds
\begin{eqnarray}\label{pre 1}\nonumber
\ib{r} |\nabla u|^p \dx &\leq&  c \Bigg[ r^{-p}\ib{2r}|u|^p \dx
+ r^{-1+\beta} \ib{2r}  |u|^2 \dx\\
&& + r^{-1} \ib{2r} |u|^3\dx +  r^{-1-\beta} \ib{2r} |u|^4\dx \Bigg]
\end{eqnarray}
for all disks $B_{2r}(x_0)$. Here, the positive constant $c$ is independent of $x_0$, $r$
and $u$. 
\item If $u$ is bounded, then it follows by choosing $\beta = 0$
\begin{equation}\label{pre 2}
\ib{r} |\nabla u|^p \dx \leq c\Big(\|u\|_{L^\infty(\rz^2)}\Big)
\Bigg[r^{-p} \ib{2r} |u|^p\dx + r^{-1}\ib{2r}|u|^2\dx \Bigg]
\end{equation}
again for all disks. In particular it holds
\begin{equation}\label{pre 3}
\ibo{R} |\nabla u|^p \dx \leq c\Big(\|u\|_{L^\infty(\rz^2)}\Big) R
\end{equation}
for radii $R \geq 1$. 

If $u_\infty\in \rz^2$ is some fixed vector, then \reff{pre 2} is also valid
for the function $\tilde{u} := u-u_\infty$ in place of $u$.

\item Suppose that 
\[
\limsup_{|x| \to \infty} |u(x)| |x|^{-\gamma} < \infty
\]
for some number $\gamma$ such that
\begin{equation}\label{pre 4}
\gamma \in \left\{ 
\begin{array}{rcl}
\mbox{$[0,1)$}\psp ,&\mbox{if}& 1 < p \leq 2 \psp ,\\[2ex]
\mbox{$[-1/2,0)$}\psp ,&\mbox{if}& p > 2\psp .
\end{array}\right.
\end{equation}
Then it holds for any $R\geq 1$
\begin{equation}\label{pre 5}
\ibo{R} |\nabla u|^p\dx \leq c R^{1+3\gamma}\psp .
\end{equation}
\end{enumerate}
\end{lemma}

\noindent{\bf Proof of Lemma \ref{pre lem 1}.} \\

\noindent \emph{Ad} $i$) \& $ii$). \\

\noindent Consider $\eta \in C^\infty_0(B_{2r}(x_0))$ such that $0 \leq \eta \leq 1$,
$\eta \equiv 1$ on $B_r(x_0)$ and $|\nabla \eta| \leq c/r$. In equation \reff{in 7} we let
$\varphi = \eta^{2l}u -w$, where the field $w$ is defined on $B_{2r}(x_0)$,
vanishing on $\partial B_{2r}(x_0)$ with the properties
\begin{eqnarray}\label{pre 6}\nonumber
\div w &=& \div (\eta^{2l}u) = \nabla \eta^{2l}\cdot u \fsp\mbox{on}\msp B_{2r}(x_0)\psp ,\\
\|\nabla w\|_{L^q(B_{2r}(x_0))} &\leq & c \|\nabla \eta^{2l}\cdot u\|_{L^q(B_{2r}(x_0))} \psp .
\end{eqnarray}
Note that \reff{pre 6} holds with the same field $w$
both for the choice $q=2$ and for the choice $q=p$ (cf.~Lemma \ref{app lem 1}). The integer $l$
will be determined later. We have
\begin{eqnarray}\label{pre 7}
\nonumber
\ib{2r} DH(\eps(u)):\eps(u) \eta^{2l}\dx &=&
- \ib{2r} DH(\eps(u)): (\nabla \eta^{2l}\otimes u) \dx\\
\nonumber &&+ \ib{2r} DH(\eps(u)): \eps(w)\dx\\
\nonumber &&- \ib{2r} u^k\partial_k u \cdot u \eta^{2l}\dx + \ib{2r} u^k\partial_k u \cdot w \dx\\
&=:& T_1 + T_2 + T_3 + T_4 \psp .
\end{eqnarray}
Young's inequality yields for any $\delta > 0$
\begin{eqnarray*}
|T_1| &\leq & c \ib{2r} |\eps(u)|^{p-1} \eta^{2l-1} |\nabla \eta| |u| \dx\\
&\leq & \delta \ib{2r} |\eps(u)|^p \eta^{(2l-1)\frac{p}{p-1}}\dx
+ c(\delta) \ib{2r} |\nabla \eta|^p |u|^p \dx \\
&\leq & \delta \ib{2r} \eta^{2l} |\eps(u)|^p\dx + c(\delta) r^{-p}
\ib{2r} |u|^p \dx \psp ,
\end{eqnarray*}
provided that we choose $l$ so large that $(2l-1) p/(p-1) \geq 2l$. For small enough $\delta$
the bound for $|T_1|$ in combination with \reff{pre 7} yields
\begin{equation}\label{pre 8}
\ib{2r} |\eps(u)|^p \eta^{2l} \dx \leq 
c\Bigg[ r^{-p} \ib{2r} |u|^p \dx + |T_2| + |T_3| + |T_4|\Bigg] \psp .
\end{equation}
Next we use \reff{pre 6} for $q=p$ and obtain by Young's inequality
\begin{eqnarray*}
|T_2| &\leq & \delta \ib{2r} |\eps(u)|^p\dx + c(\delta) \ib{2r} |\eps(w)|^p\dx \\
&\leq & \delta \ib{2r} |\eps(u)|^p \dx + c(\delta) r^{-p} \ib{2r} |u|^p \dx \psp ,
\end{eqnarray*}
thus by \reff{pre 8}
\begin{equation}\label{pre 9}
\ib{r} |\eps(u)|^p \dx \leq \delta \ib{2r} |\eps(u)|^p\dx + c(\delta) r^{-p} \ib{2r}|u|^p\dx
+ \big[|T_3| + |T_4|\big] \psp .
\end{equation}
Finally we observe using an integration by parts
\begin{equation}\label{pre 10}
|T_3| = \frac{1}{2} \Bigg| \ib{2r} u^k |u|^2 \partial_k \eta^{2l} \dx \Bigg|
\leq c r^{-1} \ib{2r} |u|^3 \dx
\end{equation}
and
\[
T_4 = - \ib{2r} u^i u^k \partial_k w^i \dx \psp ,
\]
thus
\[
|T_4| \leq \Bigg[\ib{2r} |u|^4\dx\Bigg]^{\frac{1}{2}}
\Bigg[ \ib{2r} |\nabla w|^2\dx \Bigg]^{\frac {1}{2}} \psp ,
\]
and the use of \reff{pre 6} now with the choice $q=2$ shows
\begin{eqnarray}\label{pre 11}\nonumber
|T_4| &\leq& \Bigg[\ib{2r} |u|^4\dx\Bigg]^{\frac{1}{2}}
\Bigg[ r^{-2} \ib{2r} |u|^2\dx \Bigg]^{\frac {1}{2}}\\
&=& \Bigg[r^{-1-\beta}\ib{2r} |u|^4\dx\Bigg]^{\frac{1}{2}}
\Bigg[ r^{-1+\beta} \ib{2r} |u|^2\dx \Bigg]^{\frac {1}{2}}\nonumber\\
&\leq & c r^{-1+\beta} \ib{2r} |u|^2 \dx + c r^{-1-\beta} \ib{2r} |u|^4 \dx \psp .
\end{eqnarray}
Combining \reff{pre 9} with \reff{pre 10} and \reff{pre 11} and using Lemma \ref{app lem 5}
it follows
\begin{eqnarray*}
\ib{r} |\eps(u)|^p \dx &\leq& c \Bigg[r^{-p} \ib{2r} |u|^p\dx +
r^{-1+\beta} \ib{2r} \big|u|^2 \dx\\
&&+r^{-1} \ib{2r}|u|^3\dx + r^{-1-\beta} \ib{2r} |u|^4\dx \Bigg]\psp .
\end{eqnarray*}
Applying Korn's inequality in $W^1_p(B_{2r}(x_0),\rz^2)$ (cf.~Lemma \ref{app lem 2})
we arrive at \reff{pre 1}. From \reff{pre 1} the claims \reff{pre 2} and \reff{pre 3}
immediately follow.\\

For the second statement of $ii)$ we observe that $\tilde{u} = u - u_\infty$ solves equation
\reff{in 7} with the additional term $\int u_\infty^k\partial_k \tilde{u} \cdot \varphi \dx$
and the choice $\varphi = \eta^{2l} \tilde{u} - \tilde{w}$ (with an obvious meaning of
$\tilde{w}$) leads to \reff{pre 2} for $\tilde{u}$ with the help of elementary
identities like
\[
u_\infty^k \ib{2r} \partial_k \tilde{u}^i \eta^{2l} \tilde{u}^i \dx
= - \frac{1}{2} u_\infty^k \ib{2r} |\tilde{u}|^2 \partial_k \eta^{2l} \dx \psp .
\]

\noindent \emph{Ad} $iii$).\\ 

\noindent Suppose that we have
\begin{equation}\label{pre 12}
\limsup_{|x|\to \infty} |u(x)| |x|^{-\gamma} < \infty
\end{equation}
with $\gamma$ satisfying \reff{pre 4}.\\

\emph{Case 1:  $\gamma \in [0,1)$ and $1 < p \leq 2$.} In this case \reff{pre 12} implies the growth condition
\begin{equation}\label{pre 13}
\sup_{B_R(0)} |u| \leq c R^\gamma \fsp\mbox{for all}\msp R \geq 1\psp .
\end{equation}
Quoting inequality \reff{pre 1} choosing $x_0 =0$, $r=R\geq 1$ and $\beta = \gamma$,
\reff{pre 13} gives
\[
\ibo{R} |\nabla u|^p\dx \leq c \big[R^{2-p+p\gamma} + R^{1+3\gamma}\big]\psp ,
\]
and since $2-p + p \gamma \leq 1+3\gamma$, we get \reff{pre 5}.\\

\emph{Case 2: $\gamma \in [-1/2,0)$ and $p>2$.} From \reff{pre 12} we deduce the boundedness
of $u$ together with
\begin{equation}\label{pre 14}
\sup_{R \leq |x| \leq 2R} |u| \leq R^\gamma
\end{equation}
for $R$ sufficiently large. We return to the beginning of the proof and replace $\varphi$
through the modified test-function (with $\eta$ as before and with $w^*\in \wnull{1}{q}(T_R(0),\rz^2)$
given according to Lemma \ref{app lem 1} --
again we will make use both of the choice $q=2$ and of the choice $q=p$ in this Lemma)
\[
\varphi^* = \left\{\begin{array}{ccl}
u &\mbox{on}&B_R(0) \psp ,\\
\eta^{2l} u - w^*&\mbox{on}&T_R(0)\psp ,
\end{array}\right.
\]
where we always set 
\[
T_R(x_0) := B_{2R}(x_0) - \overline{B_R(x_0)}\psp .
\] 
We have
\begin{eqnarray*}
\div w^* &=& \div (\eta^{2l}u) = \nabla \eta^{2l}\cdot u \fsp\mbox{on}\msp T_{R}(0)\psp ,\\
\|\nabla w^*\|_{L^q(T_{R}(0))} &\leq & c \|\nabla \eta^{2l}\cdot u\|_{L^q(T_{R}(0))} \psp .
\end{eqnarray*}
Note that $\iao{R} \div(\eta^{2l}\cdot u) \dx = 0$. We then obtain
a version of \reff{pre 7} with $x_0 = 0$, $w$ being replaced by $w^*$ and where
in $T_2$ and $T_4$ the integration is performed over the annulus $T_R(0)$. 
In place of \reff{pre 9} we get after specifying $c(\delta)$
\begin{equation}\label{pre 15}
\ibo{R} |\eps(u)|^p \dx \leq \delta \iao{R} |\eps(u)|^p\dx + c \delta^{1-p} R^{-p} \iao{R}|u|^p\dx
+ \big[|T_3| + |T_4|\big] \psp .
\end{equation}
For $T_3$ it holds (compare \reff{pre 10})
\[
|T_3| \leq c R^{-1} \iao{R} |u|^3 \dx
\]
and for $T_4$ we just observe
\[
|T_4| \leq  c R^{-1} \Bigg[\iao{R} |u|^4 \dx\Bigg]^{\frac{1}{2}}
\Bigg[ \iao{R} |u|^2\dx\Bigg]^{\frac{1}{2}} \psp .
\]
Thus \reff{pre 15} implies (recalling \reff{pre 14})
\begin{equation}\label{pre 16}
\ibo{R} |\eps(u)|^p \dx \leq \delta \iao{R} |\eps(u)|^p\dx 
+ c \big[\delta^{1-p} R^{2-p+p\gamma} + R^{1+3\gamma}\big] \psp .
\end{equation}
Since $u$ is bounded, we can apply \reff{pre 3} to the first term on the r.h.s.~of \reff{pre 16},
hence
\begin{equation}\label{pre 17}
\ibo{R} |\eps (u)|^p \dx \leq c \big[\delta R +  \delta^{1-p}R^{2-p+p\gamma} + c R^{1+3\gamma}\big]\psp .
\end{equation}
Suppose now that we have for some $n=0$, $1$, $2$
\begin{equation}\label{pre 18}
\ibo{R}|\eps(u)|^p\dx \leq c R^{1+n\gamma} \psp ,
\end{equation}
which by \reff{pre 3} in fact is true in the case $n=0$.
Then, instead of \reff{pre 17}, we have using assumption \reff{pre 18}
\begin{equation}\label{pre 19}
\ibo{R} |\eps(u)|^p \dx \leq c \big[\delta R^{1+n\gamma} + \delta^{1-p}R^{2-p+p\gamma} 
+c R^{1+3\gamma}\big]\psp .
\end{equation}
We choose $\delta = R^{\gamma}$ in \reff{pre 19}:
\begin{eqnarray}\label{pre 20}\nonumber
\ibo{R}|\eps(u)|^p \dx &\leq& c \big[R^{1+(n+1)\gamma} + R^{\gamma -\gamma p} R^{2-p+p\gamma}
+R^{1+3\gamma}\\
&\leq& c R^{1+(n+1) \gamma} \psp ,
\end{eqnarray}
provided that we have $(n+1) \leq 3$ (which clearly is true since we suppose $n\leq 2$ -- recall
$\gamma \leq 0$ in the case under consideration) 
and if we have in addition
\begin{equation}\label{pre 21}
\gamma + 2 -p \leq 1+ (n+1)\gamma \fsp\Leftrightarrow\fsp 1- p \leq \gamma n \psp.
\end{equation}
Note that for $\gamma \in [-1/2,0]$ and $p\geq 2$ \reff{pre 21} holds true up to the choice $n=2$
and as the final result
we obtain
\begin{equation}\label{pre 22}
\ibo{R}|\eps(u)|^p\dx \leq c R^{1+3\gamma} \psp .
\end{equation}
Applying the version of Korn' s in equality stated in Lemma \ref{app lem 2}, $iii$),
to \reff{pre 22} we obtain 
\[
\ibo{R} |\nabla u|^p \dx \leq c \big[R^{1+3\gamma} + R^{-p+2+p\gamma}\big]
\]
and thereby \reff{pre 5}
which completes the proof of Lemma \ref{pre lem 1}. \hspace*{\fill} $\Box$\\

From Lemma \ref{pre lem 1} we immediately obtain

\begin{cor}\label{pre cor 1}
Suppose that $p > 2$ and that
\[
\limsup_{|x| \to \infty} |u(x)| |x|^{-\gamma} < \infty
\]
holds for some number $\gamma < - 1/3$. Then $u$ must be identically zero.
\end{cor}

\noindent{\bf Proof of Corollary \ref{pre cor 1}.} W.l.o.g.~we may assume $\gamma \in [-1/2,-1/3)$ 
since otherwise we replace the (negative) exponent $\gamma$ through $-1/2$. But then \reff{pre 5}
yields the claim by passing to the limit $R\to \infty$. \hspace*{\fill} $\Box$\\

\section{The case $1<p<2$}\label{sub}

During this section we always assume that $u\in C^1(\rz^2,\rz^2)$ is a solenoidal
field satisfying \reff{in 7} for the choice $H(\eps) =|\eps|^p$ with exponent $p\in (1,2)$.
Note that on account of Corollary I in the paper \cite{wo} of Wolf weak solutions of \reff{in 7}
from the space $W^1_{p,\loc}(\rz^2,\rz^2)$ are of class $C^1$ if we require $p > 3/2$.\\

The proofs of Theorem \ref{theo 1} and Theorem \ref{theo 2} make extensive use
of the following preliminary result, where we let 
\[
V(\eps) := \left\{\begin{array}{ccr}
|\eps|^{\frac{p-2}{2}} &\mbox{if}& \eps \not= 0\psp ,\\
0&\mbox{if}& \eps = 0 \psp .
\end{array}\right.
\]

\begin{lemma}\label{sub lem 1} 
The velocity field $u$ is an element of the space $W^2_{p,\loc}(\rz^2,\rz^2)$ and for any disk
$B_r(x_0)$ it holds (recall  $T_r(x_0) = B_{2r}(x_0) - \overline{B_r(x_0)}$)
\begin{equation}\label{sub 1}
\ib{r} V(\eps(u))^2 |\nabla \eps(u)|^2\dx \leq c \Bigg[r^{-2} \ia{r} |\nabla u|^p\dx+
r^{-1} \ia{r}|u||\nabla u|^2\dx \Bigg] \psp ,
\end{equation}
where $c$ denotes a finite constant independent of $u$, $r$ and $x_0$.
\end{lemma}

\noindent{\bf Proof of Lemma \ref{sub lem 1}.} 
The existence of the second order weak derivatives in $L^p_{\loc}(\rz^2,\rz^2)$
has been established by Naumann \cite{na} in Theorem 2 of his paper. Actually Naumann
considers slow flows, i.e.~the convective term is neglected, but his arguments cover
the case of volume forces $f\in L^{p'}_{\loc}$, and since $u$ is a $C^1$-function, we just
put $f:= -u^k \partial_k u$.\\

For proving estimate \reff{sub 1} we benefit from the basic inequality (3.24) in Wolf's paper \cite{wo}:
let $\eta \in C^\infty_0(B_{2r}(x_0))$ such that $0 \leq \eta \leq 1$, $\eta \equiv 1$ on $B_r(x_0)$
and $|\nabla^l \eta| \leq c r^{-l}$, $l=1$, $2$. Choosing
\[
S_{ij} = \frac{\partial H}{\partial \eps_{ij}}\psp , \fsp \lambda = 0 \psp ,
\fsp \xi = \eta\psp , \fsp \tilde{f} := - u^k \partial_k u
\]
and using the symbol $\pi$ for the pressure we obtain from (3.24) in \cite{wo} (replacing $r$ by $2r$)
\begin{equation}\label{sub 2}
c(p) \ib{2r} V(\eps(u))^2 |\nabla \eps(u)|^2 \eta^2 \dx
\leq \sum_{i=1}^6 I_i 
\end{equation}
with $I_i$ defined exactly as in the above reference
and for a constant $c(p) > 0$. We have ($c$ denoting positive
constants with values varying from line to line but being independent of $x_0$ and $r$)
\begin{eqnarray}\label{sub 3}
\nonumber
|I_1| &\leq & c \ia{r} |\eps(u)|^{p-1} |\nabla u| \big[|\nabla \eta|^2 + |\nabla^2 \eta|\big] \dx\\
& \leq & c r^{-2} \ia{r} |\nabla u|^p \dx
\end{eqnarray}
and by Young's inequality (using also the estimate $|\nabla^2 u| \leq c |\nabla \eps(u)|$ and
recalling the definition of $V$)
\begin{eqnarray*}
|I_2| &\leq & c \ib{2r} |\eps(u)|^{p-1} |\nabla^2 u| \eta |\nabla \eta|\dx\\
&\leq & c \ib{2r} V(\eps(u)) |\nabla \eps(u)| \eta |\eps(u)|^{\frac{p}{2}} |\nabla \eta| \dx\\
&\leq & \delta \ib{2r} V(\eps(u))^2 |\nabla \eps(u)|^2 \eta^2 \dx
+ c(\delta) \ia{r} |\eps(u)|^p |\nabla \eta|^2 \dx \psp .
\end{eqnarray*}
Choosing $\delta$ small enough and quoting \reff{sub 3} we deduce from \reff{sub 2}
\begin{eqnarray}\label{sub 4}
\nonumber
\lefteqn{\ib{2r}V(\eps(u))^2 |\nabla \eps(u)|^2\eta^2 \dx}\\
&\leq & c \Bigg[ r^{-2} \ia{r} |\nabla u|^p\dx + |I_3+I_4| + |I_5|+|I_6|\Bigg]\psp .
\end{eqnarray}
Next we rewrite the quantity $|I_3+I_4|$ in the following form:
\[
|I_3+I_4| = \Bigg| \ib{2r} \pi \partial_k (\partial_i \eta^2\partial_k u^i)\dx \Bigg|\\
= \Bigg|\ib{2r} \pi \div \varphi \dx\Bigg| \psp ,
\]
where $\varphi^k:= \partial_i \eta^2 \partial_k u^i$. From \reff{in 4} it follows that
\[
\ib{2r} \pi \div \varphi \dx = \ib{2r} DH(\eps(u)):\eps(\varphi) \dx
+ \ib{2r} u^k \partial_k u \cdot \varphi \dx \psp ,
\]
hence
\begin{eqnarray*}
|I_3+I_4| &\leq & c \Bigg[\ib{2r} |\eps(u)|^{p-1} |\nabla \eta^2| |\nabla^2u|\dx
+ \ib{2r} |\eps(u)|^{p-1} |\nabla^2 \eta^2| |\nabla u|\dx\\
&&+ \Bigg|\ib{2r} u^k\partial_k u^i \partial_l\eta^2\partial_i u^l\dx\Bigg| \Bigg]\\
&=:& c [J_1+J_2+J_3] \psp .
\end{eqnarray*}
$J_1$ is handled in the same way as $I_2$, $J_2$ corresponds to $I_1$, thus we get from \reff{sub 4}
\begin{equation}\label{sub 5}
\ib{r} V(\eps(u))^2 |\nabla \eps(u)|^2 \dx \leq
c \Bigg[ r^{-2} \ia{r} |\nabla u|^p \dx + |I_5| + |I_6| + J_3\Bigg]\psp .
\end{equation}
We estimate $I_5$:
\[
|I_5| = \Bigg| \ib{2r} u^k \partial_k u^i \partial_l u^i\partial_l \eta^2\dx\Bigg|\\
\leq  r^{-1}\ia{r} |u| |\nabla u|^2\dx\psp .
\]
For $I_6$ it holds:
\begin{eqnarray*}
|I_6| &=& \Bigg| \ib{2r} u^k \partial_k u^i\partial_l \partial_l u^i \eta^2\dx\Bigg|
= \Bigg| \ib{2r} \partial_l (u^k\partial_k u^i \eta^2)\partial_l u^i \dx\Bigg|\\
&=& \Bigg|\ib{2r} \partial_l u^k \partial_k u^i\partial_l u^i \eta^2 \dx
+ \ib{2r} u^k \partial_l\partial_k u^i \eta^2\partial_l u^i\dx\\
&&+ \ib{2r} u^k \partial_k u^i \partial_l u^i \partial_l \eta^2 \dx\Bigg|\\
&=:& |K_1+K_2+K_3| \psp .
\end{eqnarray*}
Since we are in the $2$ D-case, we have $K_1=0$. For $K_2$ we observe
\begin{eqnarray*}
|K_2| &=& \Bigg| \ib{2r} \frac{1}{2} u^k \partial_k |\nabla u|^2 \eta^2\dx\Bigg|
= \Bigg| \ia{r} \frac{1}{2} u^k |\nabla u|^2 \partial_k \eta^2\dx\Bigg|\\
&\leq & c r^{-1} \ia{r} |u||\nabla u|^2\dx\psp ,
\end{eqnarray*}
and clearly the same bound holds for $K_3$. With \reff{sub 5} we therefore arrive at
\begin{eqnarray}\label{sub 6}\nonumber
\lefteqn{\ib{r} V(\eps(u))^2 |\nabla \eps(u)|^2\dx}\\ 
&\leq&
c \Bigg[r^{-2} \ia{r} |\nabla u|^p \dx + R^{-1} \ia{r} |u||\nabla u|^2\dx + J_3 \Bigg]\psp .
\end{eqnarray}
By the definition of $J_3$ we finally have
\[
J_3 \leq c r^{-1} \ia{r} |u| |\nabla u|^2\dx \psp ,
\]
and our claim \reff{sub 1} follows from \reff{sub 6}.\hspace*{\fill}$\Box$\\

With the help of Lemma \ref{sub lem 1} we now give the\\

\noindent{\bf Proof of Theorem \ref{theo 1}.} 
Suppose that $1<p<2$ and that we have \reff{theo 1 1} together with \reff{theo 1 2}
(the case $p=2$ together with bounded field $u$  
follows by the same arguments setting $\alpha = 0$).\\

From Lemma \ref{pre lem 1}, $iii)$, it follows with the choice $x_0=0$
on account of $\alpha < 1/3$
\begin{equation}\label{sub 7}
\lim_{R\to \infty} R^{-2} \ibo{R} |\nabla u|^p \dx = 0 \psp .
\end{equation}
Thus \reff{sub 1} will imply
\begin{equation}\label{sub 8}
V(\eps(u))^2 |\nabla \eps(u)|^2 = 0 \fsp\mbox{a.e.~on $\rz^2$}
\end{equation}
as soon as we can show that the remaining integral on the r.h.s.~of \reff{sub 1} can 
be estimated in a suitable way.\\

Obviously it is also sufficient to discuss the integral of $|u| |\nabla u|^2$ with
$T_r(x_0)$ replaced by $\Delta_r(x_0):= B_{3r/2}(x_0) -\overline{B_r(x_0)}$.
In fact, inequality \reff{sub 1} remains true with $\Delta_r(x_0)$ as domain of integration
on the r.h.s., which follows by appropriate choice of $\eta$.\\

In order to estimate the integral $\int_{\Delta_r(x_0)}|u| |\nabla u|^2\dx$ we choose
a new cut-off function $\eta \in C^\infty_0(B_{2r}(x_0))$ such that $0 \leq \eta \leq 1$,
$\eta \equiv 1$ on $\Delta_r(x_0)$ and $|\nabla \eta| \leq c/r$. 
Moreover, we note that \reff{theo 1 2} implies with a positive constant
\[
|u(x)| \leq c (1+|x|^2)^{\frac{\alpha}{2}} =: h(x)\psp .
\]
Using this bound we obtain after an integration by parts
\begin{eqnarray*}
r^{-1} \int_{\Delta_r(x_0)} |u| |\nabla u|^2\dx &\leq &
c r^{-1} \ib{2r} h \eta^2 \partial_k u^i\partial_k u^i\dx\\
&=& - c r^{-1} \ib{2r} h u^i \partial_k\partial_k u^i \eta^2\dx \\
&& - c  r^{-1} \ib{2r}  h  u^i\partial_k u^i\partial_k \eta^2 \dx\\
&& - c r^{-1} \ib{2r} \partial_k h  u^i \partial_k u^i \eta^2 \dx\\
&\leq & c r^{-1} \ib{2r} (1+|x|)^{2\alpha}  |\nabla\eps(u)|\dx \\
&& +c r^{-2} \ib{2r} (1+|x|)^{2\alpha} |\nabla u|\dx + c r^{-1}|T| \psp ,
\end{eqnarray*}
where
\[
T:= \ib{2r} \partial_k h u^i \partial_k u^i \eta^2 \dx\psp .
\]
On the set $[\eps(u)=0]$ we clearly have $\nabla \eps(u) =0$, if $\eps(u) \not = 0$, then
we use the definition of $V(\eps)$ and obtain from Young's inequality
\begin{eqnarray}\label{sub 9}
\nonumber r^{-1} \int_{\Delta_r(x_0)} |u| |\nabla u|^2\dx &\leq & 
c r^{-1} \ib{2r}  (1+|x|)^{2\alpha}  V(\eps(u)) |\nabla \eps(u)| |\eps(u)|^{1-\frac{p}{2}} \dx\\
&&+c r^{-2} \ib{2r}  (1+|x|)^{2\alpha}   |\nabla u|\dx
+ c  r^{-1} |T| \nonumber\\
&\leq & \delta \ib{2r} V(\eps(u))^2 |\nabla \eps(u)|^2\dx\nonumber\\
&&+ c(\delta) r^{-2} \ib{2r} (1+|x|)^{4\alpha}   |\eps(u)|^{2-p}\dx\nonumber\\
&&+ c r^{-2} \ib{2r}  (1+|x|)^{2\alpha}|\nabla u|\dx   
+ c  r^{-1} |T|\psp .
\end{eqnarray}
Let us look at the quantity $T$: it holds
\begin{eqnarray*}
T&=& \ib{2r} \partial_k h \frac{1}{2} \partial_k |u|^2 \eta^2 \dx\\
&=& - \ib{2r} \partial_k \partial_k h \frac{1}{2} |u|^2 \eta^2 \dx 
- \ib{2r} \partial_k h \frac{1}{2} |u|^2 \partial_k \eta^2 \dx \psp ,
\end{eqnarray*}
hence (recalling the bound for $|u|$ and the definition of $h$)
\[
|T| \leq c \Bigg[\ib{2r} (1+|x|)^{3\alpha -2}\dx + r^{-1} \ib{2r} (1+|x|)^{3\alpha -1} \dx \Bigg]\psp .
\]
It is worth remarking that the quantity $\ib{2r}h u^i\partial_k u^i\partial_k \eta^2\dx$
could have been estimated in a similar way.
We insert \reff{sub 9} combined with the estimate for $|T|$ 
into the r.h.s.~of \reff{sub 1} (in the version for the annulus
$\Delta_r(x_0)$ in place of $T_r(x_0)$) with the result
\begin{eqnarray}\label{sub 10}
\lefteqn{\ib{r} V(\eps(u))^2 |\nabla \eps(u)|^2\dx}\nonumber\\ 
&\leq & 
\delta \ib{2r} V(\eps(u))^2 |\nabla \eps(u)|^2\dx
+ c(\delta) \Bigg[r^{-2} \ib{2r}  |\nabla u|^p\dx\nonumber\\
&& +r^{-2} \ib{2r} (1+|x|)^{4\alpha} |\nabla u|^{2-p}\dx
+ r^{-2} \ib{2r}(1+|x|)^{2\alpha}  |\nabla u|\dx\nonumber\\
&&  +  r^{-1} \ib{2r} (1+|x|)^{3\alpha-2}\dx + r^{-2} \ib{2r} (1+|x|)^{3\alpha -1}\dx\Bigg]\psp .
\end{eqnarray}
Note that \reff{sub 10} holds for all $\delta >0$ and any disk $B_{2r}(x_0)$.
Then Lemma \ref{app lem 5} applied to \reff{sub 10} yields for all disks
\begin{eqnarray}\label{sub 11}
\lefteqn{\ib{r} V(\eps(u))^2|\nabla \eps(u)|^2\dx}\nonumber\\ 
&\leq & c  \Bigg[r^{-2} \ib{2r}  |\nabla u|^p\dx\nonumber\\
&& +r^{-2} \ib{2r} (1+|x|)^{4\alpha} |\nabla u|^{2-p}\dx
+ r^{-2} \ib{2r}(1+|x|)^{2\alpha}  |\nabla u|\dx\nonumber\\
&&  +  r^{-1} \ib{2r} (1+|x|)^{3\alpha-2}\dx + r^{-2} \ib{2r} (1+|x|)^{3\alpha -1}\dx\Bigg]\psp .
\end{eqnarray}
At this point we make the particular choice $x_0 = 0$. We obtain
for $r=R$ sufficiently large
\begin{eqnarray}\label{sub 12}
\lefteqn{\ibo{R} V(\eps(u))^2|\nabla \eps(u)|^2\dx}\nonumber\\
&\leq & c  \Bigg[R^{-2} \ibo{2R}  |\nabla u|^p\dx\nonumber\\
&&+R^{-2+4\alpha} \ibo{2R}  |\nabla u|^{2-p}\dx
+ R^{-2+2\alpha} \ibo{2R}  |\nabla u|\dx\nonumber\\
&& +  R^{-1} \ibo{2R} (1+|x|)^{3\alpha-2}\dx + R^{-2} \ibo{2R} (1+|x|)^{3\alpha -1}\dx\Bigg]\psp .
\end{eqnarray}
The first integral on the r.h.s.~of \reff{sub 12} is already discussed in \reff{sub 7}.
For the second one we observe with the help of \reff{pre 5}:
\begin{eqnarray*}
R^{-2+4\alpha} \ibo{2R} |\nabla u|^{2-p}\dx &\leq & 
c R^{-2+4\alpha} \Bigg[\ibo{2R} |\nabla u|^p \dx\Bigg]^{\frac{2-p}{p}}
R^{2 \frac{2p-2}{p}}\\
&= & cR^{-2+4\alpha} R^{(1+3\alpha) \frac{2-p}{p}}R^{2 \frac{2p-2}{p}}\\
&=&c R^{\frac{p-2}{p}} R^{\alpha\frac{p+6}{p}}\to 0 \fsp\mbox{as}\msp R\to \infty\psp ,
\end{eqnarray*}
where we used the fact that \reff{theo 1 1} is equivalent to
\[
\frac{p-2}{p} + \alpha \frac{p+6}{p} < 0 \psp .
\]
Next we note that \reff{theo 1 1} gives by elementary calculations
\begin{equation}\label{sub 13}
\alpha < \frac{1}{2p+3}\psp ,
\end{equation}
which shows
\begin{eqnarray*}
R^{-2+2\alpha} \ibo{2R}|\nabla u|\dx &\leq & 
c R^{-2+2\alpha} \Bigg[\ibo{2R}|\nabla u|^p\dx\Bigg]^{\frac{1}{p}} R^{2(1-\frac{1}{p})}\\
&\leq& c R^{-2+2 \alpha +\frac{1+3\alpha}{p} + 2 -\frac{2}{p}}\\ 
&= & c R^{-\frac{1}{p} + \alpha \frac{2p+3}{p}}\to 0 \fsp\mbox{as}\msp R\to \infty\psp .
\end{eqnarray*}
Finally we discuss the last two integrals on the r.h.s.~of \reff{sub 12}: 
we have
\begin{eqnarray*}
R^{-1} \ibo{2R} (1+|x|)^{3\alpha -2} \dx &=& 
2\pi R^{-1} \int_0^{2R} (1+t)^{3\alpha -2} t \dt\\
&\leq & 2\pi R^{-1} \int_0^{2R} (1+t)^{3\alpha -1} \dt\\
&=& \frac{2\pi}{3\alpha} R^{-1} \big[(1+2R)^{3\alpha}-1\big] \to 0
\end{eqnarray*}
as $R\to \infty$ on account of $\alpha < 1/3$. Moreover,
\[
R^{-2} \ibo{2R} (1+|x|)^{3\alpha -1} \dx \leq cR^{-2} R^{3\alpha -1} \to 0
\]
as $R\to \infty$, and with \reff{sub 12} we have shown
\[
\int_{\rz^2} V(\eps(u))^2 |\nabla \eps(u)|^2\dx = 0 \psp ,
\]
which implies \reff{sub 8}.\\

On the set $[\eps(u) = 0]$ we once more observe $\nabla \eps(u) =0$,
hence $\nabla^2 u =0$ by recalling the inequality $|\nabla^2 u| \leq c |\nabla \eps(u)|$ a.e.
On the set $[\eps(u) \not= 0]$ we deduce $\nabla \eps(u) = 0$ from \reff{sub 8}.
Thus $\nabla^2 u = 0$ on $\rz^2$, which means that $u$ is affine. However, since we assume
the growth condition \reff{theo 1 2}, the constancy of $u$ is established, which completes the proof of 
Theorem \ref{theo 1}. \hspace*{\fill}$\Box$\\

The proof of Theorem \ref{theo 2} additionally needs the following auxiliary results:

\begin{lemma}\label{sub lem 2}
If $u$ is as in Lemma \ref{sub lem 1}, then $v:= |\eps(u)|^{\frac{p}{2}}$ belongs to the space
$W^1_{2,\loc}(\rz^2)$ and
\[
\iom |\nabla v|^2\dx \leq c \iom V(\eps(u))^2|\nabla \eps(u)|^2\dx
\]
for any domain $\Omega \Subset \rz^2$.
\end{lemma}

\noindent {\bf Proof of Lemma \ref{sub lem 2}.} Let $v_\delta := (\delta + |\eps(u)|)^{p/2}$, $\delta > 0$.
From $u \in W^2_{p,\loc}(\rz^2,\rz^2)$ it easily follows that
$v_\delta \in W^1_{2,\loc}(\rz^2)$ together with
\begin{equation}\label{subb 1}
|\nabla v_\delta|^2 \left\{\begin{array}{clcl}
\leq & c V(\eps(u))^2 |\nabla \eps(u)|^2 &\mbox{on the set}&[\eps(u) \not= 0]\psp ,\\
= &0&\mbox{on the set}&[\eps(u) = 0] \psp ,
\end{array}\right.
\end{equation}
so that the sequence $\{v_\delta\}$ is locally uniformly bounded in $W^1_{2,\loc}(\rz^2)$,
thus 
\[
v_\delta \rightharpoondown : \tilde{v}\fsp\mbox{in}\msp W^1_{2,\loc}(\rz^2)\psp .
\]
Clearly $\tilde{v} = v$, and the desired estimate for $\iom |\nabla v|^2\dx$
follows from \reff{subb 1} and lower semicontinuity. \hspace*{\fill}$\Box$\\

\begin{lemma}\label{sub lem 3}
Suppose that $v \in C^1(\rz^2)$ satisfies $\int_{\rz^2} |\nabla v|^p \dx < \infty$
for some $p\in (1,2)$. Then it holds
\[
\limsup_{R\to \infty} R^{-2} \ibo{R} |v|\dx < \infty \psp ,
\]
in particular we deduce for any $\beta > 2$
\[
\lim_{R\to \infty} R^{-\beta} \ibo{R} |v|\dx = 0 \psp .
\]
\end{lemma}

\noindent {\bf Proof of Lemma \ref{sub lem 3}.} W.l.o.g.~let $x_0 =0$ and fix some real number $\gamma >0$.
Introducing polar coordinates $r$, $\theta$ we define
\[
f(r,\theta) = |v(r\cos(\theta), r \sin(\theta))| + \gamma \psp .
\]
The following calculations are essentially due to Gilbarg and Weinberger
(see \cite{gw}, proof of Lemma 2.1). We have by H\"older's inequality
\begin{eqnarray*}
\lefteqn{\frac{\D}{\D r} \Bigg[\int_0^{2\pi} f(r,\theta)^p \dth\Bigg]^{\frac{1}{p}}}\\
&\leq & \Bigg[\int_0^{2\pi} f(r,\theta)^p \dth\Bigg]^{\frac{1}{p}-1}
\int_0^{2\pi} f(r,\theta)^{p-1} |f_r(r,\theta)| \dth\\
&\leq & \Bigg[\int_0^{2\pi} f(r,\theta)^p \dth\Bigg]^{\frac{1}{p}-1}
\Bigg[\int_0^{2\pi} f(r,\theta)^p \dth\Bigg]^{\frac{p-1}{p}}
\Bigg[\int_0^{2\pi} |f_r(r,\theta)|^p \dth\Bigg]^{\frac{1}{p}} \psp ,
\end{eqnarray*}
where we use the symbol $f_r$ for the partial derivative of $f$ with respect
to the variable $r$. Thus, for any $\gamma > 0$ we have shown (recall that $f$ is depending on the
parameter $\gamma$)
\begin{equation}\label{subb 2}
\frac{\D}{\D r} \Bigg[\int_0^{2\pi} f(r,\theta)^p \dth\Bigg]^{\frac{1}{p}}
\leq \Bigg[\int_0^{2\pi} |f_r(r,\theta)|^p \dth\Bigg]^{\frac{1}{p}} \psp .
\end{equation}
Now let
\[
\varphi(t) := \Bigg[\int_0^{2\pi} f(t,\theta)^p \dth\Bigg]^{\frac{1}{p}} \psp .
\]
From \reff{subb 2} we get for any $R>1$:
\begin{eqnarray*}
\varphi(R) - \varphi(1) &\leq & \int_1^R \Bigg[\int_0^{2\pi} |f_r(r,\theta)|^p \dth\Bigg]^{\frac{1}{p}}\dr\\
&=&  \int_1^R \Bigg[\int_0^{2\pi} |f_r(r,\theta)|^p \dth\Bigg]^{\frac{1}{p}}r^{\frac{1}{p}}
r^{-\frac{1}{p}}\dr\\
&\leq & \Bigg[\int_1^R \Bigg[\int_0^{2\pi} |f_r(r,\theta)|^p \dth\Bigg] r\dr\Bigg]^{\frac{1}{p}}
\Bigg[ \int_1^R r^{-\frac{1}{p} \frac{p}{p-1}}\dr\Bigg]^{1-\frac{1}{p}}\psp ,
\end{eqnarray*}
where we have used H\"older's inequality once more. This shows (recall $p< 2$)
\[
\varphi(R) \leq \varphi(1) 
+ c(p)  \Bigg[\int_1^R \int_0^{2\pi} |f_r(r,\theta)|^p r \dth \dr\Bigg]^{\frac{1}{p}}
\]
and since
\[
|f_r(r,\theta)| \leq |\nabla v|(re^{i\theta}) \psp ,
\]
we deduce
\begin{equation}\label{subb 3}
\varphi(R) \leq \varphi(1) + c(p) \Bigg[ \int_{B_R(0) - \overline{B_1(0)}} |\nabla v|^p\dx\bigg]^{\frac{1}{p}} \psp .
\end{equation}
In \reff{subb 3} we pass to the limit $\gamma \to 0$ and the finiteness of the energy
then yields the inequality
\begin{equation}\label{subb 4}
\sup_{R\geq 1} \int_0^{2\pi} |v(R \cos(\theta),R \sin(\theta))|^p \dth < \infty \psp .
\end{equation}
Hence, for any $R>1$ we obtain from \reff{subb 4}
\begin{eqnarray*}
\ibo{R} |v|^p\dx &=& \int_0^R \int_0^{2\pi} |v(r\cos(\theta), r\sin(\theta))|^p r \dth\dr\\
&\leq& c +  \int_1^R \int_0^{2\pi} |v(r\cos(\theta), r\sin(\theta))|^p r \dth\dr\\
&\leq & c (1+R^2) \psp ,
\end{eqnarray*}
which proves Lemma \ref{sub lem 3}.\hspace*{\fill}$\Box$\\

\noindent {\bf Proof of Theorem \ref{theo 2}.} Now our assumption on $u$ is
\begin{equation}\label{subb 5}
\int_{\rz^2} |\nabla u|^p \dx < \infty \psp ,
\end{equation}
and in view of this hypothesis and by quoting Lemma \ref{sub lem 1} we have to discuss the quantity
\[
r^{-1} \ia{r} |u| |\nabla u|^2 \dx
\]
in order to verify \reff{sub 8} for the situation at hand. Let
\[
A := \Mint_{T_r(x_0)} u \dx \psp .
\]
Clearly it holds
\begin{equation}\label{subb 6}
r^{-1} \ia{r} |u| |\nabla u|^2 \dx \leq c r^{-1} \ia{r} |u-A| |\nabla u|^2 \dx
+c r^{-1} |A | \ia{r} |\nabla u|^2 \dx \psp .
\end{equation}
In \reff{subb 6} we apply H\"older's and Young's inequality and get for any $\delta >0$
\begin{eqnarray}\label{subb 7}
\nonumber
r^{-1} \ia{r} |u| |\nabla u|^2\dx &\leq& c \Bigg[ \ia{r} \Big[\frac{|u-A|}{r}\Big]^{\frac{p}{p-1}}\dx\Bigg]^{
\frac{p-1}{p}} \Bigg[\ia{r} |\nabla u|^{2p} \dx \Bigg]^{\frac{1}{p}}\\
&& + \delta \ia{r} |\nabla u|^{2p} \dx\nonumber\\
&& + c(\delta) r^2 \Bigg[r^{-3} \ia{r} |u|\dx\Bigg]^{\frac{p}{p-1}} \psp .
\end{eqnarray}
To the first integral on the r.h.s.~of \reff{subb 7} we apply the Sobolev-Poincar\'{e} inequality:
let $p^* := 2p'/(2+p')$, $p' := p/(p-1)$, so that $p'$ is the Sobolev exponent of $p^*$.\\

Let us first consider the case $p\geq 4/3$ for which $p^* \leq p$. Then we have
\[
\Bigg[\ia{r} |u-A|^{p'}\dx \Bigg]^{\frac{1}{p'}}
\leq c \Bigg[ \ia{r} |\nabla u|^{p^*}\dx \Bigg]^{\frac{1}{p^*}}\psp ,
\]
and by H\"older's inequality
\begin{eqnarray*}
\Bigg[\ia{r} |u-A|^{p'}\dx \Bigg]^{\frac{1}{p'}}
&\leq& c \Bigg[ \ia{r} |\nabla u|^p\dx\Bigg]^{\frac{1}{p}} r^{2(1-\frac{p^*}{p})\frac{1}{p^*}}\\
&=& c r^{3-\frac{4}{p}} \Bigg[\ia{r} |\nabla u|^p\dx\Bigg]^{\frac{1}{p}} \psp .
\end{eqnarray*}
We therefore obtain
\begin{eqnarray}\label{subb 8}
\nonumber
r^{-1} \ia{r} |u| |\nabla u|^2 \dx &\leq & c r^{2-\frac{4}{p}} \Bigg[\ia{r} |\nabla u|^p\dx\Bigg]^{\frac{1}{p}}
\Bigg[\ia{r} |\nabla u|^{2p}\dx\Bigg]^{\frac{1}{p}}\\
&& + \delta \ia{r} |\nabla u|^{2p}\dx\nonumber\\
&& + c(\delta) r^2 \Bigg[r^{-3} \ia{r} |u|\dx\Bigg]^{\frac{p}{p-1}}  \psp .
\end{eqnarray}
Let $\gamma := 2 - 4/p$ and assume w.l.o.g.~that $p < 2$, hence $\gamma < 0$.
Using our assumption \reff{subb 5} in \reff{subb 8}, we find
\begin{eqnarray*}
r^{-1} \ia{r} |u| |\nabla u|^2\dx &\leq & \delta \ia{r} |\nabla u|^{2p}\dx 
+ c r^{\gamma} \Bigg[\ia{r} |\nabla u|^{2p}\dx\Bigg]^{\frac{1}{p}}\\
&& + c(\delta) r^2 \Bigg[r^{-3} \ia{r} |u|\dx \Bigg]^{\frac{p}{p-1}} \psp ,
\end{eqnarray*}
and another application of Young's inequality shows
\begin{eqnarray}\label{subb 9}\nonumber
r^{-1} \ia{r} |u| |\nabla u|^2 \dx &\leq & 2 \delta \ia{r} |\nabla u|^{2p} \dx\\
&&+ c(\delta) \Bigg[r^{\gamma \frac{p}{p-1}} + r^2 \Bigg[r^{-3}\ia{r} |u|\dx\Bigg]^{\frac{p}{p-1}}\Bigg] \psp . 
\end{eqnarray}
Next we discuss the quantity $\ib{2r} |\nabla u|^{2p}\dx$: by Korn's inequality Lemma \ref{app lem 2}, $ii$), 
we have
\begin{equation}\label{subb 10}
\ib{2r} |\nabla u|^{2p}\dx \leq c\Bigg[\ib{2r} |\eps(u)|^{2p} \dx + r^{-2p} \ib{2r} |u|^{2p} \dx \Bigg]\psp .
\end{equation}
Since $u$ is a function of class $C^1(\rz^2,\rz^2)$ 
and thereby an element of the space $W^1_{2p,\loc}(\rz^2,\rz^2)$
we can apply the $L^{2p}$-variant of Korn's inequality
to get \reff{subb 10}. Let $B:= \Minttext_{B_{2r}(x_0)} u \dx $ and $q:= 4p/(2+2p)$, i.e.~$2p$
is the Sobolev exponent of $q$. We therefore get from the Sobolev-Poincar\'{e} inequality
\begin{eqnarray*}
\|u\|_{L^{2p}(B_{2r}(x_0))} &\leq & c \Big[ \|u-B\|_{L^{2p}(B_{2r}(x_0))}  + |B| r^{\frac{1}{p}}\Big]\\
&\leq & c \Big[\|\nabla u\|_{L^{q}(B_{2r}(x_0))} + |B| r^{\frac{1}{p}}\Big]\\
&\leq & c \Bigg[ \Bigg[\ib{2r}|\nabla u|^p \dx\Bigg]^{\frac{ 1}{p}} r^{2(\frac{1}{q}-\frac{1}{p})} 
+|B| r^{\frac{1}{p}}\Bigg] \psp ,
\end{eqnarray*}
hence (quoting \reff{subb 5})
\begin{equation}\label{subb 11}
r^{-2p} \ib{2r} |u|^{2p}\dx \leq c \big[r^{-2} + |B|^{2p} r^{2-2p}\big]\psp .
\end{equation}
By Lemma \ref{sub lem 2} the function $v:= |\eps(u)|^{p/2}$ is in the local space $W^1_{2,\loc}(\rz^2)$,
and from Lemma \ref{app lem 4} we obtain
\begin{eqnarray*}
\ib{2r} |\eps(u)|^{2p}\dx &\leq & c \Bigg[ \ib{2r} |\eps(u)|^p\dx \ib{2r} |\nabla v|^2 \dx\\
&& + r^{-2} \Bigg[ \ib{2r} |\eps(u)|^p\dx \Bigg]^2\Bigg]\psp ,
\end{eqnarray*}
thus by \reff{subb 5} and the estimate for $\ib{2r}|\nabla v|^2\dx$ stated in Lemma \ref{sub lem 2}
we find
\begin{equation}\label{subb 12}
\ib{2r} |\eps(u)|^{2p} \dx \leq c
\Bigg[ \ib{2r} V(\eps(u))^2 |\nabla \eps(u)|^2 \dx + r^{-2}\Bigg]\psp .
\end{equation}
Inserting \reff{subb 10}-\reff{subb 12} into \reff{subb 9} we get
\begin{eqnarray}\label{subb 13}
\nonumber
r^{-1} \ia{r} |u| |\nabla u|^2 \dx &\leq & 2 \delta \ib{2r} V(\eps(u))^2 |\nabla \eps(u)|^2\dx\\
&&+ c(\delta) \Bigg[ r^{-2} + |B|^{2p} r^{2-2p}
+ r^{\gamma \frac{p}{p-1}}\nonumber\\
&& + r^2 \Bigg[r^{-3} \ia{r} |u|\dx\Bigg]^{\frac{p}{p-1}}\Bigg]\psp .
\end{eqnarray}
Next we return to \reff{sub 1} estimating the second term on the r.h.s.~through \reff{subb 13} with the
result (replacing $\delta$ by $\delta/2$)
\begin{eqnarray*}
\ib{r} V(\eps(u))^2 |\nabla \eps(u)|^2 \dx &\leq &
\delta \ib{2r} V(\eps(u))^2 |\nabla \eps(u)|^2 \dx\\
&& + c(\delta) \Bigg[r^{-2} + r^{\gamma \frac{p}{p-1}}
+ r^{2-2p} \Bigg[\Mint_{B_{2r}(x_0)} |u|\dx\Bigg]^{2p}\\
&&+ r^2 \Bigg[R^{-3} \ib{2r} |u|\dx\Bigg]^{\frac{p}{p-1}} \Bigg]\psp .
\end{eqnarray*}
Applying the $\delta$-Lemma \ref{app lem 5} we arrive at (after choosing $r=R\geq 1$ and $x_0 =0$)
\begin{eqnarray}\label{subb 14}
\nonumber
\ibo{R} V(\eps(u))^2 |\nabla \eps(u)|^2\dx &\leq & c \Bigg[R^{-2} + R^{\gamma \frac{p}{p-1}}
+ \Bigg[ R^{\frac{1}{p}-3} \ibo{2R} |u|\dx\Bigg]^{2p} \\
&&+ \Bigg[ R^{2\frac{p-1}{p} -3} \ibo{2R}|u|\dx\Bigg]^{\frac{p}{p-1}} \Bigg]\psp .
\end{eqnarray}
By Lemma \ref{sub lem 3} it follows that the r.h.s.~of \reff{subb 14} vanishes as $R\to \infty$, thus we obtain
\reff{sub 8} and, as outlined at the end of the proof of Theorem \ref{theo 1},
$u$ has to be an affine function. But then \reff{subb 5} yields the constancy of $u$, which proves Theorem
\ref{theo 2} in the case $p \geq 4/3$.\\

If $6/5 < p < 4/3$ we return to \reff{subb 8} and estimate the r.h.s.~of the inequality
stated in \reff{subb 7} in a different way: observing that by the choice of $p$
\[
p < p^* = \frac{2p}{3p-2} < 2 p \psp ,
\]
we can apply the interpolation inequality
\[
\|\nabla u\|_{p^*} \leq \|\nabla u\|_p^\alpha \|\nabla u\|_{2p}^{1-\alpha} \psp ,
\]
where all norms are calculated over $T_r(x_0)$ and where
\[
\frac{1}{p^*} = \frac{\alpha}{p} + \frac{1-\alpha}{2p}\psp , \fsp\mbox{hence}
\fsp \alpha = \frac{2p}{p^*} - 1 \psp .
\]
This gives using \reff{subb 5}
\begin{eqnarray*}
\Bigg[ \ia{r} \Big|\frac{u-A}{r}\Big|^{\frac{p}{p-1}} \dx\Bigg]^{\frac{p-1}{p}}
\Bigg[ \ia{r} |\nabla u|^{2p} \dx\Bigg]^{\frac{1}{p}}
&\leq & c r^{-1} \|\nabla u\|_{p^*} \|\nabla u\|^2_{2p}\\
&\leq & c r^{-1}\|\nabla u\|^\alpha_p \|\nabla u\|_{2p}^{2+1-\alpha}\\
& \leq & c r^{-1} \Bigg[\ia{r}|\nabla u|^{2p}\dx\Bigg]^{\frac{3-\alpha}{2p}}\psp .
\end{eqnarray*}
With elementary calculations one obtains
\[
\frac{3-\alpha}{2p} = \frac{6-3p}{2p}
\]
and we find that
\[
\frac{3-\alpha}{2p} < 1
\]
is true under our hypothesis $p > 6/5$. This gives us the flexibility to apply Young's inequality
with the result
\[
r^{-1}\Bigg[ \ia{r} |\nabla u|^{2p} \dx \Bigg]^{\frac{3-\alpha}{2p}}
\leq c \Bigg[r^{-\kappa} + \ia{r} |\nabla u|^{2p}\dx\Bigg]
\]
with a suitable positive exponent $\kappa$. Using this estimate in \reff{subb 7} the proof can be
finished as before.\hspace*{\fill}$\Box$\\

\section{The case $p >2$}\label{sup}
We start with an appropriate variant of Lemma \ref{sub lem 1} which is more difficult
to establish since now we can no longer benefit from the higher weak differentiability results
of Naumann \cite{na} and Wolf \cite{wo}.

\begin{lemma}\label{sup lem 1}
Let $u \in C^1(\rz^2,\rz^2)$ denote a solenoidal field satisfying \reff{in 7} with $H(\eps) = |\eps|^p$
for some exponent $p > 2$. Moreover, let
\[
W:= W(\eps(u)) := |\eps(u)|^{\frac{p-2}{2}} \eps(u) \psp .
\]
Then it holds:
\begin{enumerate}
\item $W$ is in the space $W^1_{2,\loc}(\rz^2,\rz^{2\times 2})$.
\item There exists a finite constant $c$ independent of $u$ such that for any $\delta > 0$
and for each $q > 2$
\begin{eqnarray}\label{sup 1}\nonumber
\ib{r} |\nabla W|^2 \dx & \leq &\delta
\ib{2r} |\nabla W|^2 \dx + c\Bigg[\delta^{-1} r^{-2} \ia{r} |\nabla u|^p\dx \\
&& + r^{-1} \Bigg[\ia{r} |u|^{\frac{q}{q-2}} \dx\Bigg]^{1-\frac{2}{q}} 
\Bigg[\ia{r} |\nabla u|^q\dx\Bigg]^{\frac{2}{q}}\Bigg] 
\end{eqnarray}
for any disk $B_r(x_0)$.
\end{enumerate}
\end{lemma}

\noindent\emph{Proof.} We use the difference quotient technique and let
\[
\da v(x) := \frac{1}{h} \big(v(x+he_\alpha)-v(x)\big)
\]
for functions $v$, parameters $h\not= 0$ and a coordinate direction $e_\alpha$, $\alpha = 1$, $2$.
If $\varphi \in C^1_0(\rz^2,\rz^2)$ satisfies $\div \varphi = 0$, then we have the equation\reff{in 7}
together with the identity
\[
0 = \int_{\rz^2} DH(\eps(u))(x+he_\alpha):\eps(\varphi)(x)\dx
+\int_{\rz^2} (u^k \partial_k u^i)(x+he_\alpha) \varphi^i(x)\dx \psp ,
\]
hence after subtracting the equations and after dividing by $h$
\begin{equation}\label{sup 2}
\int_{\rz^2} \da\big(DH(\eps(u))\big):\eps(\varphi)\dx + \int_{\rz^2}\da (u^k \partial_k u)
\cdot \varphi \dx = 0 \psp ,
\end{equation}
and \reff{sup 2} clearly extends to solenoidal fields from $W^1_{p,\loc}(\rz^2,\rz^2)$
with compact support. Alternatively -- taking into account the pressure function $\pi$ in 
the weak form of \reff{in 4} -- we can replace \reff{sup 2} by
\begin{eqnarray}\label{sup 3}
\nonumber
0&=& \int_{\rz^2} \da \big(DH(\eps(u))\big):\eps(\varphi)\dx
+ \int_{\rz^2} \da (u^k \partial_k u) \cdot \varphi \dx
 - \int_{\rz^2} \da \pi \div \varphi \dx\nonumber\\
&=:& T_1 + T_2 + T_3 
\end{eqnarray}
valid for all $\varphi \in W^1_{p,\loc}(\rz^2,\rz^2)$ with compact support.
In \reff{sup 3} we choose $\varphi := \varphi_\alpha := \eta^2 \da u$
with $\alpha =1$, $2$ being fixed (no summation convention w.r.t.~$\alpha$) and with
$\eta \in C^2_0(B_{2r}(x_0))$, $0\leq \eta \leq 1$, $\eta = 1$ on $B_r(x_0)$,
$|\nabla \eta| \leq c r^{-1}$. We discuss the quantities $T_i$ from \reff{sup 3}
related to our choice of $\varphi$: it holds
\begin{eqnarray*}
T_1 &=& \ib{2r} \da \big(DH(\eps(u))\big) :\eps(\da u) \eta^2 \dx\\
&&+ \ib{2r} \da \big( DH(\eps(u))\big) : (\nabla \eta^2\otimes \da u)\dx\\
&=:& U_1 + U_2\psp ,
\end{eqnarray*}
and for $U_1$ we observe
\begin{eqnarray*}
\lefteqn{\da \big(|\eps(u)|^{p-2}\eps(u)\big)(x):\eps(\da u)(x)}\\
&=& \frac{1}{h} \Big[|\eps(u)|^{p-2}(x+he_\alpha)\eps(u)(x+he_\alpha)
-|\eps(u)|^{p-2}(x)\eps(u)(x)\Big]:\\
&&\qquad\frac{1}{h} \Big[\eps(u)(x+he_\alpha) - \eps(u)(x)\Big]\\
&\geq& c \Big[|\eps(u)|^{p-2}(x+h e_\alpha) + |\eps(u)|^{p-2}(x)\Big]\da \eps(u)(x): \da\eps(u)(x) \psp ,
\end{eqnarray*}
where the last inequality can be easily deduced from Lemma \ref{app lem 6}, $ii$).
At the same time, Lemma \ref{app lem 6}, $i$), implies
\begin{eqnarray*}
\lefteqn{ \frac{1}{|h|} \Big| |\eps(u)|^{p-2}(x+he_\alpha)\eps(u)(x+he_\alpha)
-|\eps(u)|^{p-2}(x)\eps(u)(x)\Big|}\\
&\leq & c \Big[ |\eps(u)|^2(x+he_\alpha) + |\eps(u)|^2(x)\Big]^{\frac{p-2}{2}}
\frac{1}{|h|} \big| \eps(u)(x+he_\alpha) - \eps(u)(x)\big|\psp ,
\end{eqnarray*}
thus using Young's inequality
\begin{eqnarray*}
|U_2| &\leq & c \ib{2r} \big[|\eps(u)|(x+he_\alpha) + |\eps(u)|(x)\big]^{p-2} |\da\eps(u)| |\da u| \nabla \eta|^2 \dx\\
&\leq & \delta \ib{2r} \big[|\eps(u)|(x+he_\alpha) + |\eps(u)|(x)\big]^{p-2} \da \eps(u) :\da \eps(u) \eta^2 \dx\\
&&+ c \delta^{-1} \ib{2r} \big[|\eps(u)|(x+he_\alpha) + |\eps(u)|(x)\big]^{p-2} |\nabla \eta|^2 \da u \cdot \da u \dx
\end{eqnarray*}
for any $\delta > 0$. Combining these estimates, returning to \reff{sup 3} and choosing $\delta$ small
enough we find
\begin{eqnarray}\label{sup 4}
\lefteqn{ \ib{2r} \Big[|\eps(u)|^{p-2}(x+he_\alpha) + |\eps(u)|^{p-2}(x)\Big] \eta^2 \da \eps(u):\da \eps(u)\dx}
\nonumber\\
&\leq & c \Bigg[\ia{r} \Big[ |\eps(u)|^{p-2}(x+he_\alpha) + |\eps(u)|^{p-2}(x)\Big] |\nabla \eta|^2 \da u \cdot \da u \dx
\nonumber\\
&& + |T_2| + |T_3|\Bigg]\psp .
\end{eqnarray}
Next we look at the pressure term $T_3$: we have
\[
\div (\eta^2 \da u) = \nabla \eta^2 \cdot \da u =: f_h^\alpha
\]
where the function $f_h^\alpha$ is compactly supported in $T_r(x_0)$. 
Moreover, we have by the definition of $f^\alpha_h$ and the properties
of $\eta$
\begin{eqnarray*}
\ia{r} f^\alpha_h \dx &=& \ia{r} \div(\eta^2 \cdot \da u) \dx\\
&=& - \int_{\partial B_r(x_0)} \da u(x) \cdot \frac{x-x_0}{r} \psp\D{\cal H}^1(x)\\
&=& - \ib{r} \div (\da u)\dx = 0 \psp ,
\end{eqnarray*}
where ${\cal H}^1$ denotes the one-dimensional Hausdorff-measure.
According to Lemma \ref{app lem 1} we find $\psi^\alpha_h \in \wnull{1}{p}(T_r(x_0),\rz^2)$
satisfying $\div \psi^\alpha_h = f^\alpha_h$ on
$T_r(x_0)$ and sharing the usual estimates on the annulus $T_r(x_0)$. We get
\begin{eqnarray*}
|T_3| &=& \Bigg| \ia{r} \da \pi \div(\eta^2 \da u) \dx \Bigg|
= \Bigg| \ia{r} \da \pi f^\alpha_h \dx \Bigg|\\
&=& \Bigg| \ia{r} \da \pi \div \psi^\alpha_h \dx \Bigg|
\end{eqnarray*}
and if we use \reff{sup 3} with $\psi^\alpha_h$ as test function it follows
\begin{eqnarray}\label{sup 5}
|T_3| &=& \Bigg| \ia{r} \da \big(DH(\eps(u))\big) : \eps(\psi^\alpha_h)\dx
+ \ia{r} \da (u^k \partial_k u) \cdot \psi^\alpha_h \dx \Bigg|\nonumber\\
&=:& |S_1 + S_2|\psp .
\end{eqnarray}
For $S_1$ we first observe (compare the discussion of $U_2$)
\begin{eqnarray*}
|S_1| & \leq & c \ia{r} \big| \da (|\eps(u)|^{p-2} \eps(u))\big| |\eps(\psi^\alpha_h)|\dx\\
&\leq & c \ia{r} \big(|\eps(u)|(x+he_\alpha) + |\eps(u)|(x)\big)^{p-2} |\da \eps(u)| |\eps(\psi^\alpha_h)|\dx
\end{eqnarray*}
and then use Young's inequality to get for any $\delta > 0$
\begin{eqnarray}\label{sup 6}
|S_1| & \leq & \delta \ia{r} \big(|\eps(u)| (x+he_\alpha) + |\eps(u)|(x)\big)^{p-2} |\da \eps(u)|^2\dx\nonumber\\
&& + c \delta^{-1} \ia{r} \big(|\eps(u)|(x+he_\alpha) + |\eps(u)|(x)\big)^{p-2}
|\eps(\psi^\alpha_h)|^2\dx \psp .
\end{eqnarray}
According to \cite{ga}, Theorem 3.2, p.~130, the support of $\psi^\alpha_h$ is compact in $T_r(x_0)$
and by quoting Lemma 7.23 of \cite{gt} we can estimate using H\"older's inequality
\begin{eqnarray}\label{sup 7}
\lefteqn{
c \delta^{-1} \ia{r} \big(|\eps(u)|(x+he_\alpha) + |\eps(u)|(x)\big)^{p-2} |\eps(\psi^\alpha_h)|^2\dx}\nonumber\\
&\leq & c \delta^{-1} \Bigg[\ia{r} |\eps(\psi^\alpha_h)|^p \dx\Bigg]^{\frac{2}{p}}
\Bigg[ \ia{r} |\nabla u|^p \dx\Bigg]^{1-\frac{2}{p}}\nonumber\\
&\leq & c \delta^{-1} r^{-2} \ia{r} |\nabla u|^p\dx \psp .
\end{eqnarray}
We apply a similar reasoning to the first term on the r.h.s.~of \reff{sup 4} and get
from \reff{sup 4}--\reff{sup 7}
\begin{eqnarray}\label{sup 8}
\lefteqn{\ib{r} \eta^2 \big(|\eps(u)|^{p-2}(x+he_\alpha)+ |\eps(u)|^{p-2}(x)\big)
|\da \eps(u)|^2\dx}\nonumber\\
&\leq & \delta \ia{r} \big(|\eps(u)|^{p-2} (x + h e_\alpha) + |\eps(u)|^{p-2}(x)\big) |\da \eps(u)|^2|\dx\nonumber\\
&& + c \delta^{-1} r^{-2} \ia{r} |\nabla u|^p \dx + c \big[|T_2| + |S_2|\big]
\end{eqnarray}
with $T_2$ defined in \reff{sup 3} for the choice
$\varphi = \eta^2 \da u$ and $S_2$ from \reff{sup 5}. Let us look at $T_2$: we have
\begin{eqnarray*}
T_2 &=& \ib{2r} \da (u^k \partial_k u^i) \eta^2 \da u^i \dx\\
&=& \ib{2r} \da u^k \partial_k u^i \da u^i \eta^2 \dx
+ \ib{2r} u^k \partial_k (\da u^i) \da u^i \eta^2 \dx\\
&=& \ib{2r} \da u^k \partial_k u^i \da u^i \eta^2 \dx
- \frac{1}{2} \ib{2r} u^k \big(\da u \cdot \da u\big) \partial_k \eta^2 \dx \psp ,
\end{eqnarray*}
hence
\begin{equation}\label{sup 9}
|T_2| \leq c \Bigg[ \ib{2r} \big(\da u \cdot \da u\big)|\nabla u|\dx
+ \frac{1}{r} \ia{r} \big(\da u \cdot \da u\big) |u| \dx \Bigg]\psp .
\end{equation}
For estimating $S_2$ we again use the properties of $\psi^\alpha_h$ as already done
after \reff{sup 6}:
\begin{eqnarray*}
S_2 &=& - \ia{r} u^k \partial_k u \cdot \Delta^\alpha_{-h} \psi^\alpha_h \dx\\
&\leq & \Bigg[ \ia{r} |\nabla \psi^\alpha_h|^2\dx\Bigg]^{\frac{1}{2}}
\Bigg[ \ia{r} |u|^2 |\nabla u|^2\dx \Bigg]^{\frac{1}{2}}\\
&\leq & c r^{-1} \Bigg[ \ia{r} |\nabla u|^2\dx \Bigg]^{\frac{1}{2}}
\Bigg[ \ia{r} |u|^2 |\nabla u|^2\dx \Bigg]^{\frac{1}{2}} \psp ,
\end{eqnarray*}
thus
\begin{equation}\label{sup 10}
|S_2| \leq c \Bigg[ r^{-1} \ia{r} |\nabla u|^2 \dx + r^{-1} \ia{r} |u|^2 |\nabla u|^2\dx \Bigg]\psp .
\end{equation}
Inserting \reff{sup 9} and \reff{sup 10} into \reff{sup 8} and using the $\delta$-Lemma
\ref{app lem 5} with suitable functions $f$, $f_j$ and $g$ 
(replacing the domain of integration $T_r(x_0)$ through $B_{2r}(x_0)$ on the r.h.s.~of the
inequalities under consideration), we deduce
\begin{equation}\label{sup 11}
\ib{r} \big(|\eps(u)|^{p-2}(x+he_\alpha) + |\eps(u)|^{p-2}(x)\big)
|\da \eps(u)|^2 \dx \leq c(r,u) < \infty
\end{equation}
for a constant $c(r,u)$ being independent of $h$. Now it is easy to see (cf.~Lemma \ref{app lem 6}, $i$))
that
\[
\da W(\eps(u)):\da W(\eps(u))
\]
can be bounded from above by the quantity
\[
\big(|\eps(u)|^{p-2}(x+he_\alpha) + |\eps(u)|^{p-2}(x)\big) |\da \eps(u)|^2\psp ,
\]
so that \reff{sup 11} implies
\begin{equation}\label{sup 12}
W(\eps(u)) \in W^1_{2,\loc}(\rz^2,\rz^{2\times 2}) \psp .
\end{equation}
At the same time we can deduce from \reff{sup 8} and the subsequent estimates by taking from
now on the sum w.r.t.~$\alpha$ (letting $W=W(\eps(u))$ and using
the formulas for $T_2$, $S_2$)
\begin{eqnarray}\label{sup 13}
\lefteqn{\ib{r} \da W : \da W\dx}\nonumber\\
&\leq & \delta \ib{2r} \da W: \da W \dx + c \Bigg[\delta^{-1} r^{-2} \ia{r} |\nabla u|^p \dx\nonumber\\
&&+ \ib{2r} \big|\da u^k \partial_k u^i \da u^i\big|\dx + r^{-1} \ia{r} |u| \big(\da u \cdot \da u\big)\dx\nonumber\\
&& + r^{-1} \ia{r} |u| |\nabla u|^2 \dx + r \ia{r} |u| | \Delta^\alpha_{-h} \psi^\alpha_h|^2 \dx \Bigg]\psp .
\end{eqnarray}
Here the third and the fourth integral on the r.h.s.~correspond to $T_2$, whereas
the last two ones are produced by breaking up $S_2$ with the help of Young's inequality.
Using the properties of $\psi^\alpha_h$ we can estimate the last integral on the r.h.s.~of \reff{sup 13}
by H\"older's inequality in order to get for any $q >2$
\begin{eqnarray*}
\ia{r} |u| |\Delta^\alpha_{-h} \psi^\alpha_h|^2 \dx &\leq &
 \Bigg[ \ia{r} |u|^{\frac{q}{q-2}}\Bigg]^{1-\frac{2}{q}} 
\Bigg[\ia{r} |\Delta^\alpha_{-h} \psi^\alpha_h|^q\dx\Bigg]^{\frac{2}{q}}\\
&\leq & c r^{-2}\Bigg[ \ia{r} |u|^{\frac{q}{q-2}}\Bigg]^{1-\frac{2}{q}}
\Bigg[\ia{r} |\nabla u|^q \dx\Bigg]^{\frac{2}{q}}\psp ,
\end{eqnarray*}
If we insert this estimate into \reff{sup 13}, we obtain after passing to the limit $h\to 0$
(using $\partial_\alpha u^k \partial_k u^i \partial_\alpha u^i\equiv 0$)
\begin{eqnarray}\label{sup 14}
\ib{r} |\nabla W(\eps(u))|^2\dx &\leq &
\delta \ib{2r} |\nabla W(\eps(u))|^2 \dx \nonumber\\
&&+ c \Bigg[ \delta^{-1} r^{-2} \ia{r} |\nabla u|^p\dx
+ r^{-1} \ia{r} |u| |\nabla u|^2 \dx\nonumber\\
&&+r^{-1} \Bigg[\ia{r} |u|^{\frac{q}{q-2}}\dx\Bigg]^{1-\frac{2}{q}}
\Bigg[\ia{r} |\nabla u|^q\dx\Bigg]^{\frac{2}{q}}\Bigg]\psp ,
\end{eqnarray}
and \reff{sup 14} holds for all $\delta > 0$, all disks $B_r(x_0)$ and for any $q>2$.
Hence, with \reff{sup 14} our claim \reff{sup 1} is established.\hspace*{\fill}$\Box$\\

We also need a substitute for Lemma \ref{sub lem 3}.
\begin{lemma}\label{sup lem 2}
Suppose that $v \in C^1(\rz^2)$ satisfies $\int_{\rz^2} |\nabla v|^p \dx < \infty$ for some $p\in (2,\infty)$. Then we have
\[
\limsup_{R \to \infty} \frac{1}{R^{3-\frac{2}{p}}} \ibo{R} |v| \dx < \infty \psp .
\]
\end{lemma}

\noindent {\bf Proof of Lemma \ref{sup lem 2}.} From the proof of Lemma \ref{sub lem 3} we recall the inequality
\[
\varphi(R) - \varphi(1) \leq \Bigg[\int_1^R \int_0^{2\pi} |f_r(r,\theta)|^p\dth r\dr\Bigg]^{\frac{1}{p}}
\Bigg[\int_1^R r^{-\frac{1}{p} \frac{p}{p-1}}\dr\Bigg]^{1-\frac{1}{p}}
\]
being valid also for $p\geq 2$. In place of \reff{subb 3} we obtain (recalling $|f_r(r,\theta)| \leq |\nabla v(r e^{i\theta})|$)
\[
\varphi(R) \leq \varphi(1) + c(p) R^{\frac{p-2}{p}} \Bigg[\int_{B_R(0)-B_1(0)} |\nabla u|^p\dx \Bigg]^{\frac{1}{p}} \psp ,
\]
provided we choose $R \geq 1$. Using the finiteness of the energy we get after passing to the limit $\gamma \to 0$
\[
\sup_{R \geq 1} R^{2-p} \int_0^{2\pi} |v(R\cos(\theta),R\sin(\theta))|^p \dth <  \infty \psp .
\]
This estimate implies for $R \geq 1$
\begin{eqnarray*}
\ibo{R} |v|^p \dx &=& \int_0^R \int_0^{2\pi} |v(r\cos(\theta), r\sin(\theta))|^p r \dth \dr\\
&\leq & c + \int_1^R \int_0^{2\pi} |v(r\cos(\theta),r\sin(\theta))|^p r \dth \dr\\
&\leq &c(1+R^p) \leq c R^p \psp .
\end{eqnarray*}
Finally we make use of H\"older's inequality
\[
\ibo{R} |v|\dx \leq c \Bigg[ \ibo{R} |v|^p \dx \Bigg]^{\frac{1}{p}} R^{2 (1-\frac{1}{p})} \psp ,
\]
hence our claim follows by inserting the previous estimate. \hspace*{\fill}$\Box$\\

Next we give the\\

\noindent {\bf Proof of Theorem \ref{theo 4}.} W.l.o.g.~let $u_\infty = 0$.
Let us further assume that
\begin{equation}\label{th4 1}
\sup_{|x| \geq R} |u(x)| |x|^{-\gamma} \to 0 \fsp \mbox{as}\msp R\to \infty
\end{equation}
for some $\gamma \in [-1/3,0)$, hence we have for all $R \geq 1$:
\begin{equation}\label{th4 2}
|u(x)|\leq \Theta(R) R^\gamma \fsp\mbox{for all}\msp R\leq |x| \leq 2R
\end{equation}
with some function $\Theta$ such that $\Theta(R) \to 0$ as $R \to \infty$.
From \reff{sup 1} we deduce choosing $q=p$ and applying Young's inequality ($W:= W(\eps(u))$)
\begin{eqnarray*}
\ib{r} |\nabla W|^2\dx &\leq &\delta \ib{2r} |\nabla W|^2 \dx
+ c \Bigg[\delta^{-1} r^{-2} \ib{2r} |\nabla u|^p \dx\\
&& + r^{-1} \Bigg[\ib{2r} |u|^{\frac{p}{p-2}}\dx \Bigg]^{1-\frac{2}{p}}
\Bigg[\ib{2r} |\nabla u|^p\dx\Bigg]^{\frac{2}{p}}\Bigg]\\
& \leq & \delta \ib{2r} |\nabla W|^2 \dx
+ c \Bigg[\delta^{-1} r^{-2} \ib{2r} |\nabla u|^p \dx\\
&& + r^{-1}\Bigg[ \tau \ib{2r} |\nabla u|^{p}\dx
+ \tau^{-\frac{2}{p-2}} \ib{2r} |u|^{\frac{p}{p-2}}\dx\Bigg]\Bigg]
\end{eqnarray*}
for any disk $B_r(x_0)$. Let $\tau := r^{\para}$
for some $\para \in (0,1)$. The $\delta$-Lemma \ref{app lem 5} yields for any disk $B_r(x_0)$
\begin{eqnarray}\label{th4 3}\nonumber
\ib{r} |\nabla W|^2\dx &\leq & c \Bigg[r^{-2} \ib{2r} |\nabla u|^p\dx
+ r^{-1+\para}\ib{2r} |\nabla u|^p \dx\\
&& +r^{-\frac{2\para}{p-2}-1} \ib{2r} |u|^{\frac{p}{p-2}}\dx \Bigg] \psp .
\end{eqnarray}
We choose $x_0=0$, $r=R >1$ and insert \reff{pre 5} in \reff{th4 3}, 
where the last integral on the r.h.s.~of \reff{th4 3} is handled with
the condition $|u| \leq c$. We arrive at
\begin{eqnarray*}
\ibo{R} |\nabla W|^2\dx &\leq &c \big[R^{-2+1+3\gamma} + R^{-1+\para + 1 + 3\gamma}
+R^{-\frac{2 \para}{p-2}-1}R^2\big]\\
&\leq & c \big[ R^{\para + 3 \gamma} + R^{1-\frac{2\para}{p-2}}\big]\psp ,
\end{eqnarray*}
i.e.~we have with some $\nu < 1$ (w.l.o.g.~$\nu > 0$)
\begin{equation}\label{th4 4}
\ibo{R} |\nabla W|^2 \dx \leq c R^\nu\fsp\mbox{for all}\msp R \geq 1\psp .
\end{equation}
Next we choose $\mu \in (\nu,1)$ and apply \reff{sup 1} with $q=p$ and $\delta = R^{-\mu}$ to obtain
\begin{eqnarray}\label{th4 5}
\ibo{R} |\nabla W|^2 \dx &\leq & c \Big[R^{-\mu + \nu} + R^{\mu -2 +1 + 3 \gamma}\nonumber \\
&&+R^{-1} R^{2-\frac{4}{p}} \sup_{R \leq |x| \leq 2R} |u| R^{(1+3\gamma)\frac{2}{p}}\Big]\psp .
\end{eqnarray}
By the choice of the above parameters, the first two terms on the r.h.s.~of \reff{th4 5} converge
to zero as $R\to \infty$ and it remains to discuss the quantity (recall \reff{th4 2})
\[
\zeta_R := R^{1-\frac{4}{p}} \Theta(R) R^\gamma R^{(1+3\gamma)\frac{2}{p}}
= \Theta(R) R^{1-\frac{2}{p}+\gamma(1+\frac{6}{p})} \psp ,
\]
where we have to distinguish the three different cases of Theorem \ref{theo 4}.\\

\emph{Case 1.} For $2 < p < 6$ we may choose $\gamma = (2-p)/(p+6)$ in
\reff{th4 1}, where we note that
\[
\gamma > - \frac{1}{3} \fsp\Leftrightarrow\fsp p < 6 \psp .
\]
This particular choice of $\gamma$ gives
\[
1 - \frac{2}{p} + \gamma \big(1+\frac{6}{p}\big) = 0
\]
which implies $\zeta_R \to 0$ as $R \to \infty$, hence the first part of the theorem is established.\\

\emph{Case 2.} For $p=6$ we have by assumption
\[
|u(x)| \leq c R^{-\frac{1}{3}} \fsp\mbox{for all}\msp |x| \geq R
\]
and for all $R \geq 1$.
Since the condition $\Theta(R) \to 0$ as $R \to \infty$ is not needed for deriving \reff{th4 4},
we obtain \reff{th4 4} as before. Moreover, \reff{pre 5} gives
\begin{equation}\label{th4 6}
\int_{\rz^2} |\nabla u|^p \dx < \infty \psp .
\end{equation}
As above we let $q=p$ and $\delta = R^{-\mu}$ in \reff{sup 1} to obtain (recall \reff{th4 4})
\begin{equation}\label{th4 7}
\ibo{R} |\nabla W|^2 \dx \leq c \Bigg[R^{\nu -\mu} + R^{\mu -2}
+ R^{-1} \Bigg[\iao{R} |u|^{\frac{3}{2}}\dx\Bigg]^{\frac{2}{3}}
\Bigg[\iao{R} |\nabla u|^6 \dx\Bigg]^{\frac{1}{3}} \Bigg]\psp .
\end{equation}
Here we observe
\[
R^{-1} \Bigg[\iao{R} |u|^{\frac{3}{2}}\dx\Bigg]^{\frac{2}{3}}
\leq c R^{-1} R^{-\frac{1}{3}} R^{2 \frac{2}{3}} \leq c
\]
and by \reff{th4 6} the last integral of \reff{th4 7} converges to $0$ as $R\to \infty$ which
completes the proof in the second case of Theorem \ref{theo 4}.\\

\emph{Case 3.} In the case $p > 6$ we again have by assumption 
the global energy estimate \reff{th4 6}.
We recall \reff{pre 15} of Section \ref{pre}, choose $\delta = 1/2$ 
in this inequality
and observe that by the boundedness of $u$
\[
R^{-p} \iao{R} |u|^p \dx \to 0 \fsp\mbox{as}\msp R \to \infty \psp .
\]
Moreover we have
\[
|T_3| + |T_4| \leq c R \Big[\sup_{R \leq |x| \leq 2R} |u|\Big]^{3} \to 0 \fsp\mbox{as}\msp R \to \infty\psp .
\]
As a consequence we see
\[
\int_{\rz^2} |\eps(u)|^p \dx \leq \frac{1}{2} \int_{\rz^2} |\eps(u)|^p \dx
\]
which means $\eps(u)\equiv 0$, hence $u$ is a rigid motion and $u = const$ by the decay
assumption. This completes the proof of Theorem \ref{theo 4}.\hspace*{\fill}$\Box$\\

We finish this section with the\\ 

\noindent{\bf Proof of Theorem \ref{theo 3}.} Let $2 < p \leq 3$.  
As above we have \reff{th4 3}, where we know in
the situation at hand that
\[
\int_{\rz^2} |\nabla u|^p \dx < \infty \psp ,
\]
hence for any $R \geq 1$ ($W:= W(\eps(u))$)
\begin{equation}\label{th3 1}
\ibo{R} |\nabla W|^2 \dx \leq c \Bigg[R^{-1+\para} + R^{-\frac{2\para}{p-2}-1} 
\ibo{2R} |u|^{\frac{p}{p-2}}\dx \Bigg] \psp .
\end{equation}
We insert \reff{th3 1} in the r.h.s.~of \reff{sup 1} choosing $q=p$ there and get for any $\delta >0$
\begin{eqnarray}\label{th3 2}
\ibo{R} |\nabla W|^2 \dx &\leq & \delta \Bigg[ R^{-1+\para} + R^{-\frac{2\para}{p-2}-1}
\ibo{2R} |u|^{\frac{p}{p-2}}\dx \Bigg]\nonumber\\
&& + c \Bigg[\delta^{-1} R^{-2} \iao{R} |\nabla u|^p \dx\nonumber\\
&& + R^{-1} \Bigg[\iao{R} |u|^{\frac{p}{p-2}}\dx\Bigg]^{\frac{p-2}{p}}
\Bigg[\iao{R} |\nabla u|^p\dx \Bigg]^{\frac{2}{p}} \Bigg]\psp .
\end{eqnarray}
Let 
\[
A:= \Mint_{B_{2R}(0)} u \dx
\]
and observe
\begin{eqnarray}\label{th3 3}
\ibo{2R} |u|^{\frac{p}{p-2}} \dx &\leq & c \Bigg[\ibo{2R} |u-A|^{\frac{p}{p-2}} \dx
+ R^2 |A|^{\frac{p}{p-2}}\Bigg]\nonumber\\
&\leq & c \Bigg[ \ibo{2R} |u - A|^{\frac{p}{p-2}}\dx + \Bigg|R^{-2 + 2 \frac{p-2}{p}}
\ibo{2R}u \dx\Bigg|^{\frac{p}{p-2}}\Bigg]\psp .
\end{eqnarray}
To the first integral on the r.h.s.~of \reff{th3 3} we apply the Sobolev-Poincar\'{e} inequality,
which is possible on account of $p/(p-2) > 2$: letting
\[
1 < q := \frac{2p}{3p-4}
\]
and observing $q < p$ on account of $p>2$, we find
\begin{eqnarray}\label{th3 4}
\Bigg[\ibo{2R} |u-A|^{\frac{p}{p-2}}\dx \Bigg]^{\frac{p-2}{p}} & \leq & 
c \Bigg[ \ibo{2R} |\nabla u|^q\dx \Bigg]^{\frac{1}{q}}\nonumber\\
&\leq & c \Bigg[ \Bigg[ \ibo{2R} |\nabla u|^p \dx\Bigg]^{\frac{q}{p}} R^{2(1-\frac{q}{p})}\Bigg]^{\frac{1}{q}}\nonumber\\
&=& c R^{\frac{2}{q} - \frac{2}{p}} \Bigg[\ibo{2R} |\nabla u|^p \dx\Bigg]^{\frac{1}{p}}\psp ,
\end{eqnarray}
where we also made use of H\"older's inequality. With \reff{th3 3} and \reff{th3 4} we find
\begin{eqnarray}\label{th3 5}
\xi_1 &:=& R^{-1} \Bigg[ \iao{R} |u|^{\frac{p}{p-2}}\dx\Bigg]^{\frac{p-2}{p}} \Bigg[\iao{R} |\nabla u|^p\dx\Bigg]^{\frac{2}{p}}
\nonumber\\
&\leq & R^{-1} \Bigg[ \ibo{2R} |u|^{\frac{p}{p-2}} \dx \Bigg]^{\frac{p-2}{p}}
\Bigg[\iao{R} |\nabla u|^p\dx\Bigg]^{\frac{2}{p}}\nonumber\\
&\leq & c \Bigg[R^{-1} R^{\frac{2}{q}-\frac{2}{p}} \Bigg[\iao{R} |\nabla u|^p \dx\Bigg]^{\frac{2}{p}}
\Bigg[\ibo{2R} |\nabla u|^p\dx\Bigg]^{\frac{1}{p}}\nonumber\\
&&+ R^{-1} R^{-2+ 2 \frac{p-2}{p}} \Bigg| \ibo{2R}u\dx\Bigg| \Bigg[\iao{R}|\nabla u|^p\dx\Bigg]^{\frac{2}{p}}\Bigg]\nonumber\\
&=& c \Bigg[R^{2-\frac{6}{p}} \Bigg[\iao{R} |\nabla u|^p \dx\Bigg]^{\frac{2}{p}} \Bigg[\ibo{2R} |\nabla u|^p\dx\Bigg]^{\frac{1}{p}}
\nonumber\\
&& + \Bigg| R^{-1-\frac{4}{p}} \ibo{2R} u \dx\Bigg| \Bigg[ \iao{R} |\nabla u|^p \dx\Bigg]^{\frac{2}{p}}\Bigg]\psp ,
\end{eqnarray}
and since 
\[
\lim_{R\to \infty} \iao{R} |\nabla u|^p \dx = 0
\]
it follows
\begin{equation}\label{th3 6}
\lim_{R \to \infty} \xi_1 = 0 
\end{equation}
on account of $p \leq 3$ and by quoting Lemma \ref{sup lem 2}.
Using \reff{th3 3} and \reff{th3 4} one more time we obtain
\begin{eqnarray}\label{th3 7}
\xi_2 &:=& \delta R^{-\frac{2 \para}{p-2} -1} \ibo{2R} |u|^{\frac{p}{p-2}}\dx\nonumber\\
&\leq & c \delta R^{-\frac{2\para}{p-2} -1} \Bigg[ R^{(\frac{2}{q}-\frac{2}{p}) \frac{p}{p-2}}
\Bigg[ \ibo{2R} |\nabla u|^p \dx \Bigg]^{\frac{1}{p-2}}
 + \Bigg|R^{-2 + 2 \frac{p-2}{p}} \ibo{2R} u \dx \Bigg|^{\frac{p}{p-2}} \Bigg]\nonumber\\
&=& c \delta R^{- \frac{2\para}{p-2}-1} \Bigg[R^3 \Bigg[\ibo{2R}|\nabla u|^p\dx \Bigg]^{\frac{1}{p-2}}
+ \Bigg| R^{-\frac{4}{p}} \ibo{2R} u \dx \Bigg|^{\frac{p}{p-2}} \Bigg] \psp .
\end{eqnarray}
Since $p \leq 3$, it holds
\[
- \frac{2\para}{p-2} -1+3 = 2 - \frac{2\para}{p-2} \leq 2 - 2 \para \psp .
\]
Recalling that $\para \in (0,1)$ is arbitrary, we may fix, e.g., $\para = 3/4$, hence $2 -2 \para = 1/2$.
Finally we choose $\delta = 1/R$ in \reff{th3 2}. This implies
\[
\delta R^{-\frac{2\para}{p-2}-1} R^3 \Bigg[\ibo{2R} |\nabla u|^p \dx \Bigg]^{\frac{1}{p-2}} \to 0
\]
as $R\to \infty$ and at the same time by Lemma \ref{sup lem 2}
\[
\delta R^{-\frac{2\para}{p-2}-1} \Bigg| R^{-\frac{4}{p}} \ibo{2R} u \dx \Bigg|^{\frac{p}{p-2}}
= \Bigg| R^{-2-\frac{2\para}{p}} \ibo{2R} u \dx\Bigg|^{\frac{p}{p-2}} \to 0
\]
as $R\to \infty$, hence
\begin{equation}\label{th3 8}
\lim_{R\to \infty} \xi_2 = 0 \psp .
\end{equation}
Inserting \reff{th3 5}--\reff{th3 8} into \reff{th3 2} and passing to the limit $R\to \infty$, we have shown that
$\nabla W = 0$ on $\rz^2$, hence $u$ is affine
and the finiteness of the $p$-energy implies the constancy of $u$. \hspace*{\fill}$\Box$\\

\section{Proof of Theorem \ref{theo 5}}\label{quad}

Let $u$ denote an entire solution of \reff{in 1} satisfying \reff{theo 5 1}. Introducing the vorticity
\[
\omega:= \partial_2 u^1 - \partial_1 u^2
\]
we have for $q$, $l \in \mathbb{N}$ sufficiently large with $\eta \in C^\infty_0(\rz^2)$
\begin{eqnarray}\label{th5 1}
\int_{\rz^2} \omega^{2q} \eta^{2l} \dx &=&
\int_{\rz^2} (\partial_2 u^1 - \partial_1 u^2) \omega^{2q-1}\eta^{2l}\dx\nonumber\\
&=&\int_{\rz^2} \div (-u^2,u^1) \omega^{2q-1} \eta^{2l} \dx\nonumber\\
&=& - \int_{\rz^2} (-u^2,u^1) \cdot \nabla \big[\omega^{2q-1}\eta^{2l}\big]\dx\nonumber\\
&=& (2q-1) \int_{\rz^2} \nabla \omega \cdot (u^2,-u^1) \omega^{2q-2}\eta^{2l}\dx\nonumber\\
&& + 2l \int_{\rz^2} (u^2,-u^1) \cdot \nabla \eta \omega^{2q-1} \eta^{2l-1} \dx \psp ,
\end{eqnarray}
and from $\div u =0$ we infer
\begin{eqnarray}\label{th5 2}
\int_{\rz^2} u \cdot \nabla \omega \omega^{2q-3} \eta^{2l} \dx &=&
\frac{1}{2q-2} \int_{\rz^2} u \cdot \nabla \omega^{2q-2} \eta^{2l}\dx\nonumber\\
&=& - \frac{1}{2q-2} \int_{\rz^2} u \cdot \nabla \eta^{2l} \omega^{2q-2}\dx \psp .
\end{eqnarray}
Recall that
\[
\Delta \omega - u \cdot \nabla \omega = 0 \fsp\mbox{on}\msp \rz^2\psp ,
\]
hence
\[
\int_{\rz^2} \nabla \omega \cdot \nabla \varphi \dx
+ \int_{\rz^2} u \cdot \nabla \omega \varphi \dx = 0
\]
for $\varphi \in C^1_0(\rz^2)$. We specify $\varphi = \eta^{2l} \omega^{2q-3}$ and get
\begin{eqnarray}\label{th5 3}
\lefteqn{\int_{\rz^2} \eta^{2l} (2q-3) |\nabla \omega|^2 \omega^{2q-4}\dx}\nonumber\\
&=& - \int_{\rz^2} \nabla \omega \cdot \nabla \eta^{2l} \omega^{2q-3} \dx
- \int_{\rz^2} u \cdot \nabla \omega \omega^{2q-3} \eta^{2l}\dx \psp .
\end{eqnarray}
By Young's inequality, the first term on the r.h.s.~of \reff{th5 3} is estimated through
\[
\delta \int_{\rz^2} |\nabla \omega|^2 \omega^{2q-4} \eta^{2l} \dx +
c(\delta,l) \int_{\rz^2} |\nabla \eta|^2 \eta^{2l-2} \omega^{2q-2} \dx \psp ,
\]
to the second term on the r.h.s.~of \reff{th5 3} we apply \reff{th5 2}. This yields
after appropriate choice of $\delta$
\begin{eqnarray}\label{th5 4}
\lefteqn{ \int_{\rz^2} |\nabla \omega|^2 \omega^{2q-4} \eta^{2l}\dx}\nonumber\\
&\leq & c(l,q) \Bigg[ \int_{\rz^2} \omega^{2q-2} \eta^{2l-2} |\nabla \eta|^2 \dx
+ \int_{\rz^2} |u| |\nabla \eta^{2l}| \omega^{2q-2} \dx \Bigg]\psp .
\end{eqnarray}
Now we return to \reff{th5 1} and estimate
\begin{eqnarray*}
\int_{\rz^2} \omega^{2q} \eta^{2l}\dx &\leq &
(2q-1) \int_{\rz^2}  |\nabla \omega| |u| \omega^{2q-2} \eta^{2l} \dx
+ 2l \int_{\rz^2} |u| |\nabla \eta| \omega^{2q-1} \eta^{2l-1} \dx\\
&\leq & \delta \int_{\rz^2} \omega^{2q} \eta^{2l} \dx
+ c(\delta,q) \int_{\rz^2} |\nabla \omega|^2 |u|^2 \omega^{2q-4} \eta^{2l}\dx\\
&& + 2l \int_{\rz^2} |u| |\nabla \eta| \omega^{2q-1} \eta^{2l-1}\dx \psp ,
\end{eqnarray*}
hence for $\delta$ sufficiently small
\begin{eqnarray}\label{th5 5}
\lefteqn{\int_{\rz^2} \eta^{2l} \omega^{2q} \dx}\nonumber\\
&\leq & c(l,q) \Bigg[\int_{\rz^2} |\nabla \omega|^2 |u|^2 \omega^{2q-4}\eta^{2l} \dx
+ \int_{\rz^2} |u| |\nabla \eta| \omega^{2q-1} \eta^{2l-1}\dx \Bigg]\psp .
\end{eqnarray}
Next we specify $\eta$: let $R\geq 1$ and choose $\eta =1$ on $B_R(0)$,
$0 \leq \eta \leq 1$, $\spt \eta \subset B_{2R}(0)$, $|\nabla \eta| \leq c/R$.
From \reff{theo 5 1} we get (w.l.o.g.~we assume $\alpha >0$)
\begin{equation}\label{th5 6}
|u(x)| \leq c R^\alpha\fsp\mbox{for all}\msp x \in B_R(0) \psp .
\end{equation}
We use \reff{th5 6} on the r.h.s.~of \reff{th5 5} and get
\begin{eqnarray*}
\lefteqn{\ibo{2R} \eta^{2l} \omega^{2q} \dx}\\
&\leq & c(l,q) \Bigg[R^{2\alpha} \ibo{2R} |\nabla \omega|^2 \omega^{2q-4} \eta^{2l}\dx
+ R^{\alpha} \ibo{2R} |\nabla \eta| \omega^{2q-1} \eta^{2l-1} \dx\Bigg]\psp , 
\end{eqnarray*}
and if we apply \reff{th5 4} on the r.h.s.~quoting \reff{th5 6} one more time it follows
\begin{eqnarray}\label{th5 7}
\lefteqn{\ibo{2R} \eta^{2l} \omega^{2q} \dx}\nonumber\\
& \leq & c(l,q) \Bigg[R^{2\alpha} \ibo{2R} \omega^{2q-2} \eta^{2l-2} |\nabla \eta|^2\dx
+ R^{3\alpha} \ibo{2R} |\nabla \eta^{2l}| \omega^{2q-2} \dx\nonumber\\
&& + R^\alpha \ibo{2R} \omega^{2q-1} |\nabla \eta| \eta^{2l-1} \dx\Bigg]\nonumber\\
&=:& c(l,q) \big[T_1 + T_2 + T_3\big]\psp .
\end{eqnarray}
Young's inequality yields
\begin{eqnarray*}
T_1 &\leq & \ibo{2R} \omega^{2q-2} \eta^{2l-2} R^{2\alpha -2}\dx\\
&\leq & \delta \ibo{2R} \omega^{2q} \eta^{(2l-2)2q/(2q-2)}\dx
+ c(\delta) R^{2+q(2\alpha -2)}\psp 
\end{eqnarray*}
and
\begin{eqnarray*}
T_2 &\leq & \ibo{2R} \omega^{2q-2} \eta^{2l-1} R^{3\alpha -1}\dx\\
&\leq & \delta \ibo{2R} \omega^{2q} \eta^{(2l-1)2q/(2q-2)}\dx + c(\delta) R^{2+q(3\alpha -1)} 
\end{eqnarray*}
as well as
\begin{eqnarray*}
T_3 &\leq & c \ibo{2R} \omega^{2q-1} \eta^{2l-1} R^{\alpha -1} \dx\\
&\leq & \delta \ibo{2R} \omega^{2q} \eta^{(2l-1)2q/(2q-1)} \dx + c(\delta) R^{2+2q(\alpha -1)} \psp .
\end{eqnarray*}
Moreover, for $l \gg 1$ we have
\[
2l \leq \frac{(2l-2)2q}{2q-2} \fsp\mbox{and}\fsp
2 l \leq \frac{(2l-1)2q}{2q-1} \psp ,
\]
hence, for $\delta$ small enough, we obtain from \reff{th5 7}
\begin{equation}\label{th5 8}
\ibo{2R} \eta^{2l} \omega^{2q} \dx \leq c(l,q)
\Big[R^{2+q(2\alpha -2)} + R^{2+q(3\alpha -1)} + R^{2+2q(\alpha -1)}\Big]\psp .
\end{equation}
Recall that $\alpha < 1/3$. Therefore we can fix a sufficiently large exponent $q$ with the property that
\[
2+ q (3\alpha -1) < 0 \psp ,
\]
and \reff{th5 8} shows
\[
\ibo{R} \omega^{2q}\dx \leq c(l,q) R^{2+q (3\alpha -1)} \to 0 \fsp\mbox{as}\msp R \to 0 \psp ,
\]
hence $\omega =0$ on $\rz^2$. This together with $\div u =0$ shows that $u$ is harmonic
and the constancy of $u$ then follows from \reff{theo 5 1} and results concerning
entire harmonic functions. \hspace*{\fill}$\Box$\\

\renewcommand{\thesection}{Appendix. Helpful tools}
\section{}\label{app}
\renewcommand{\thesection}{A}
\renewcommand{\theequation}{\mbox{A.\arabic{equation}}}
The following lemma is a well known result. A proof together with further comments can be found
in \cite{ga}, Chapter III, Section 3. Our formulation is taken from \cite{am}, Lemma 2.5.

\begin{lemma}\label{app lem 1}
Suppose that we are given numbers $1 < p_1 \leq p \leq p_2 < \infty$.

Then there exists a constant $c = c(p_1,p_2)$ as follows: if $f \in L^p(B_r(x_0))$ satisfies
$\Minttext_{B_r(x_0))} f \dx = 0$, then there exists a field $v$ in the space
$\wnulltext{1}{p}(B_r(x_0),\rz^2)$ satisfying $\div v =f$ on the disk $B_r(x_0)$ together with the estimate
\[
\ib{r} |\nabla v|^s \dx \leq c \ib{r} |f|^s \dx
\]
for any exponent $s \in [p_1,p]$. The same is true if the disk is replaced
by the annulus $T_r(x_0) = B_{2r}(x_0) - \overline{B_r(x_0)}$.
\end{lemma}

Our next tool is a collection of Korn-type inequalities. We refer the reader to Lemma 3.0.1 in \cite{fs},
where a list of references is given. We note that the last statement follows from the first one
by applying $i$) to $\eta v$, where $\eta$ is a suitable cut-off function. 
\begin{lemma}\label{app lem 2}
Let $1 < p < \infty$. Then there exists a constant $c(p)$ such that the following 
inequalities hold.
\begin{enumerate}
\item For all $v\in \wnull{1}{p}(B_r(x_0),\rz^2)$ we have
\[
\|\nabla v\|_{L^p(B_r(x_0))} \leq c(p) \|\eps(v)\|_{L^p(B_r(x_0))} \psp .
\]
\item For all $v\in W^{1}_{p}(B_r(x_0),\rz^2)$ we have
\[
\|\nabla v\|_{L^p(B_r(x_0))} \leq c(p) \Big[\|\eps(v)\|_{L^p(B_r(x_0))} 
+ r^{-1} \|v\|_{L^p(B_r(x_0))}\Big]\psp .
\]
\item For all $v\in W^{1}_{p}(B_{2r}(x_0),\rz^2)$ we have letting
$T_r(x_0) = B_{2r}(x_0)-\overline{B_r(x_0)}$
\[
\|\nabla v\|_{L^p(B_r(x_0))} \leq c(p) \Big[\|\eps(v)\|_{L^p(B_{2r}(x_0))} 
+ r^{-1} \|v\|_{L^p(T_r(x_0))}\Big]\psp .
\]
\end{enumerate}
\end{lemma}


The following lemma originates from the work of Ladyzhenskaya (see \cite{la}, Lemma 1, p.~8).
Actually it is a local variant of Ladyzhenskaya's lemma established as Lemma 2.6 in part B of
\cite{zh}.

\begin{lemma}\label{app lem 4}
Suppose that $u \in W^1_2(B_r(x_0))$, $B_r(x_0) \subset \rz^2$. Then there is a constant
$c$ independent of $u$, $x_0$ and $r$ such that
\[
\ib{r} |u|^4 \dx \leq c \Bigg[\ib{r} |u|^2 \dx \ib{r}|\nabla u|^2 \dx
+ r^{-2} \Bigg[  \ib{r} |u|^2\dx\Bigg]^2\Bigg]\psp .
\]
\end{lemma}

The next lemma goes back to Giaquinta and Modica (see \cite{gm}, Lemma 0.5).
We state a small extension presented in \cite{fzha} as Lemma 3.1.

\begin{lemma}\label{app lem 5}
Let $f$, $f_1$, \dots , $f_l$ denote non-negative functions from the space 
$L^1_{\loc}(\rz^2)$. Suppose further that we are given exponents $\alpha_1$, \dots , $\alpha_l > 0$.

Then we can find a number $\delta_0 > 0$ (depending on $\alpha_1$, \dots, $\alpha_l$)
as follows:
if for $\delta \in (0,\delta_0)$ it is possible to calculate a constant $c(\delta)>0$
such that the inequality
\begin{equation}\label{gm 1}
\ib{r}f\dx\leq  \delta \ib{2r} f \dx
+c(\delta) 
\sum_{j=1}^l r^{-\alpha_j}\ib{2r}f_j\dx 
\end{equation}
holds for any choice of $B_r(x_0) \subset \rz^2$,
then there is a constant $c$ with the property
\begin{equation}\label{gm 2}
\ib{r} f \dx \leq c  \sum_{j=1}^l r^{-\alpha_j} \ib{2r}f_j\dx
\end{equation}
for all disks $B_r(x_0) \subset \rz^2$.
\end{lemma}

Finally we recall some well known inequalities.

\begin{lemma}\label{app lem 6}
Let $p > 2$.
\begin{enumerate}
\item With suitable positive constants $c_1 < c_2$ it holds
\[
c_1 \big[ |\xi|^{p-2} + |\eta|^{p-2}\big] |\xi - \eta|^2 \leq 
\Big| |\xi|^{\frac{p-2}{2}} \xi - |\eta|^{\frac{p-2}{2}} \eta\Big|^2
\leq c_2 \Big[|\xi|^{p-2} + |\eta|^{p-2}\big]|\xi -\eta|^2
\]
for any $\xi$, $\eta \in \rz^M$, $M \geq 1$.
\item There exists a constant $c>0$ such that
\[
\big(|\xi|^{p-2}\xi - |\eta|^{p-2}\eta\big):(\xi - \eta) \geq c \big[|\xi|^{p-2} + |\eta|^{p-2}\big] |\xi - \eta|^2
\]
for all $\xi$, $\eta \in \rz^M$, $M \geq 1$.
\end{enumerate}
\end{lemma}

\noindent{\bf Proof.} $i$) follows from inequality (2.4) in \cite{gmrem} by letting $\mu =0$, $\delta = p-2$
in this reference.

For proving $ii$) we let $F(\xi) =|\xi|^{p-2}\xi$ and observe that
\begin{eqnarray*}
\big(F(\xi) - F(\eta)\big): (\xi - \eta) 
&=& \int_0^1 \frac{\D}{\dt} F(\eta + t(\xi - \eta))\dt:(\xi - \eta)\\
&=:& \int_0^1 |\eta + t (\xi - \eta)|^{p-2}\dt |\xi - \eta|^2 + A \psp ,
\end{eqnarray*}
where $A$ is easily seen to be non-negative. From Lemma 2.2 in \cite{fh}
we therefore deduce
\[
\big(F(\xi)-F(\eta)\big) : (\xi - \eta) \geq c |\xi - \eta|^2 \big[|\xi - \eta|^{p-2} + |\eta|^p\big]\psp ,
\]
and our claim immediately follows from this estimate by considering the cases
$|\xi| \geq 2 |\eta|$ and $|\xi| < 2 |\eta|$, respectively. \hspace*{\fill}$\Box$\\ 



\vspace{3ex}

\begin{minipage}{8cm}
Michael Bildhauer, Martin Fuchs\\
Saarland University\\
Department of Mathematics\\
P.O.~Box 15 11 50\\
D-66041 Saarbr\"ucken,\\ 
Germany\\
e-mail:\\ bibi@math.uni-sb.de\\ fuchs@math.uni-sb.de
\end{minipage}\hspace*{\fill}
\begin{minipage}[h]{8cm}
Guo Zhang\\
University of Jyv\"askyl\"a\\
Department of Mathematics and Statistics\\
P.O.~Box 35 (MaD)\\
FI.-40014\\
Finland\\
e-mail: guo.g.zhang@jyu.fi
\end{minipage}

\end{document}